\documentclass[twocolumn,aps,prl,amsmath,superscriptaddress,notitlepage,longbibliography]{revtex4-1}
\usepackage{amssymb}
\usepackage{mathrsfs}
\usepackage{graphicx}
\usepackage{float}
\usepackage[caption=false]{subfig}
\usepackage{scalerel}
\usepackage{epstopdf}
\usepackage[normalem]{ulem}
\usepackage{verbatim}
\usepackage{xcolor}
\usepackage{bm}
\usepackage{dcolumn}% Align table columns on decimal point
\usepackage{threeparttable}

\usepackage[utf8x]{inputenc}

\def\blue{\textcolor{blue}}
\def\red{\textcolor{red}}

\begin{document}

\def\qv{\vec{q}}
\def\blue{\textcolor{blue}}
\def\magenta{\textcolor{magenta}}
\def\apricot{\textcolor{Apricot}}

\def\GJ{\textcolor{black}}
\def\TT{\textcolor{ForestGreen}}
\definecolor{ora}{rgb}{1,0.45,0.2}
%{0.2, 0.7, 0.2}
\def\LH{\textcolor{black}}

\newcommand{\norm}[1]{\left\lVert#1\right\rVert}
\newcommand{\ad}[1]{\text{ad}_{S_{#1}(t)}}

\title{Exceptional Boundary States and negative Entanglement Entropy}

\author{Ching Hua Lee}
\email{phylch@nus.edu.sg}
\affiliation{Department of Physics, National University of Singapore, Singapore 117542}

%\pacs{73.43.Lp, 71.10.Pm}
\date{\today}

\date{\today}
\begin{abstract}
This work introduces a new class of robust states known as Exceptional Boundary (EB) states, which are distinct from the well-known topological and non-Hermitian skin boundary states. EB states occur in the presence of exceptional points, which are non-Hermitian critical points where eigenstates coalesce and fail to span the Hilbert space. This eigenspace defectiveness not only limits the accessibility of state information, but also interplays with long-range order to give rise to singular propagators only possible in non-Hermitian settings. Their resultant EB eigenstates are characterized by robust anomalously large or negative occupation probabilities, unlike ordinary Fermi sea states whose probabilities lie between zero and one. EB states remain robust after a variety of quantum quenches and give rise to enigmatic negative entanglement entropy contributions. Through suitable perturbations, the coefficient of the logarithmic entanglement entropy scaling can be continuously tuned. EB states represent a new avenue for robustness arising from geometric defectiveness, independent of topological protection or non-reciprocal pumping.
\end{abstract}

\maketitle

\noindent{\textit{Introduction. --}} Robust boundary states are vivid physical manifestations of deep physics. Slightly over a decade ago, they came into the spotlight as topologically protected edge states~\cite{kane2005quantum,konig2007quantum,qi2008topological,chen2009experimental,hasan2010colloquium,qi2011topological,wang2011topological,lindner2011floquet,mellnik2014spin}. More recently, much attention has also focused on non-Hermitian skin states, which are robust boundary states arising from net unbalanced gain/loss~\cite{Lee2016nonH,xiong2018does,yao2018edge,Lee2019anatomy,Lee2019hybrid,kunst2018biorthogonal,jin2019bulk,kunst2019non,edvardsson2019non,yokomizo2019non,lee2018tidal,song2019non,song2019realspace,li2019geometric,borgnia2020nonH,zhang2019correspondence,yang2019auxiliary,brandenbourger2019non,lee2019unraveling,mu2019emergent,longhi2019probing,luo2020skin,li2020critical,lee2020ultrafast,longhi2020non,lee2020many,cao2020non,xue2020non,liu2020helical,yoshida2020mirror,rosa2020dynamics,yi2020non,xiao2020non,li2020topological,schomerus2020nonreciprocal,PhysRevLett.124.086801,koch2020bulk,arouca2020unconventional,teo2020topological,okuma2020quantum,kawabata2020higher}. This work shall introduce a third and fundamentally distinct type of robust boundary state known as ``Exceptional Boundary'' (EB) states.

EB states are predicted to exist in critical non-Hermitian fermionic systems containing exceptional points (EPs)~\cite{Bender1998nonH,bender2007making,NHbook,gong2018topological,kawabata2019symmetry}. Unlike ordinary gapless points, EPs are also branch points of the complex energy Riemann surface~\cite{berry2004EP,dembowski2004encircling,Rotter2009non,jin2009solutions,longhi2010pt,heiss2001chirality,heiss2012EP,xu2016topological,Hassan2017EP,Hu2017EP,shen2018topological,wang2019arbitrary,ghatak2019new,miri2019exceptional,zhang2020non,jin2019hybrid,kawabata2019classification}, and host unusual quantum critical, fractal and thermodynamic behavior as well as potential applications in sensing, multi-mode laser manipulation and topological energy transfer~\cite{longhi2010pt,heiss2001chirality,lee2012geometric,brody2013information,wang2016wave,chen2017exceptional,mortensen2018fluctuations,lau2018fundamental,insinga2018quantum,naghiloo2019quantum,ozdemir2019parity,dora2019kibble,wiersig2020prospects,wang2020petermann,arouca2020unconventional,li2020hamiltonian}. Exceptional Boundary states around EPs are physically distinguished by their anomalously large occupation probabilities, which can be detected long after their host EPs are quenched. In the quantum entanglement context, EBs states also give rise to decreased or even negative entanglement entropy (EE), very different from what is possible in Hermitian systems.

%\red{ How to experimentally measure occupation and characteristic shape and density? Describe this only in the intro!} 

%\red{is a quench valid? What is the interpretation of that $P$? and $P^2$? Interpretation of eigenvalues outside $[0,1]$?}
%show that sufficiently singular EPs not only possess very different entanglement behavior compared to Hermitian critical points, 
Fundamentally, EB state phenomena arise because the system is not just critical, but also \emph{defective} around the exceptional point. A generic occupied state cannot be completely expanded in terms of the eigenstates, blurring the distinction between occupied and unoccupied states and resulting in an asymmetrically singular propagator. This gives rise to pronounced EB eigenstate accumulation when an effective boundary is imposed through a quench or entanglement cut. Such intriguing EB states are well-separated from the other eigenstates and possess distinctive spatial profiles, violating the celebrated notion of bulk-boundary correspondence in an unique way. They are completely distinct from other types of robust boundary eigenstates [Fig.~1], being protected by the defectiveness of the eigenspace, which is a geometric rather than dynamical or topological property,  %Interestingly, they are characterized by very large or negative effective occupation probabilities, a consequence of the extreme asymmetry around the singularity of an EP. As such, they can present dominant negative contributions to the entanglement entropy at an EP, which gives rise to anomalous or even negative values of the effective central charge. 
Perturbations that increase the effective rank at the EP induce a gradual crossover from negative EE scaling to ordinary critical scaling behavior, allowing for continuous tuning of logarithmic critical scaling coefficients. 
%Protected by defectiveness which limits the accessible state information, EB states are robust against various perturbations, although their EE gradually crosses over from negative scaling to  as perturbations are increased.
%from negative to suppress its divergence, %d even though they are no longer strictly robust, such that their effective central charges become continuously tunable.

\begin{figure}
\includegraphics[width=1\linewidth]{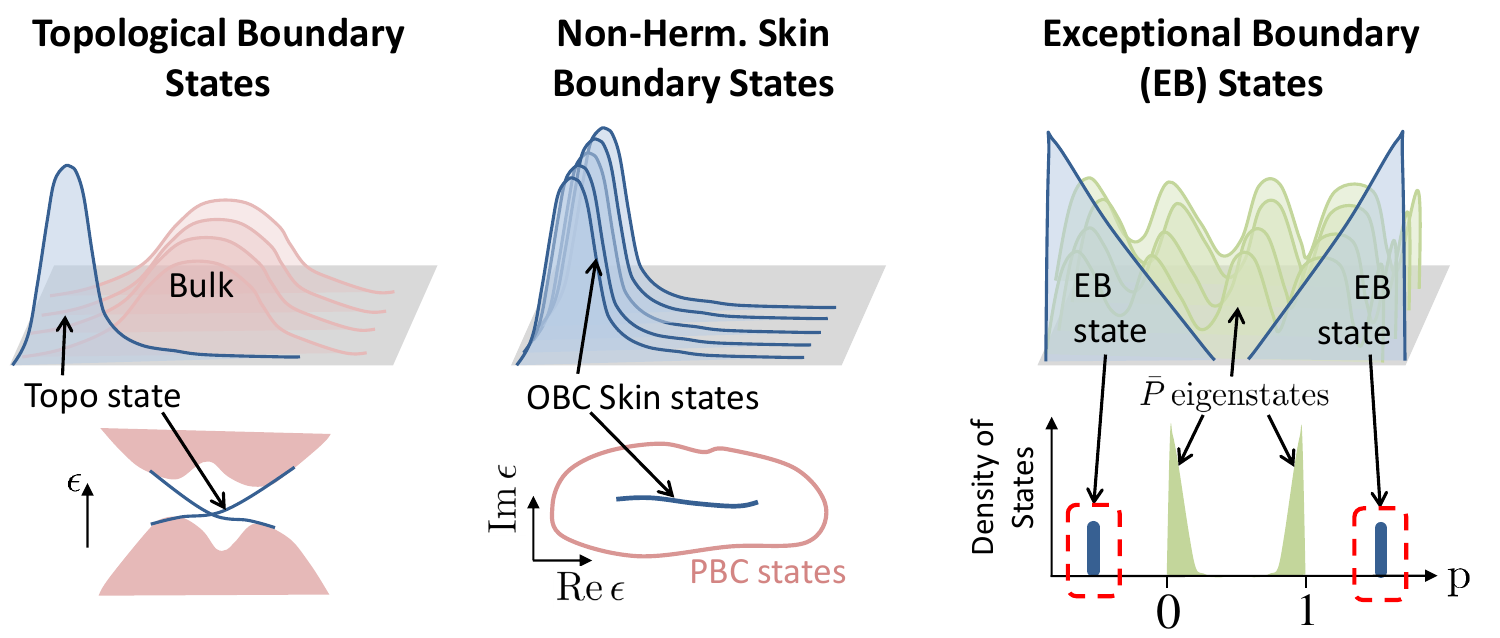}
\caption{Exceptional Boundary (EB) states as a third distinct type of robust boundary state. Left) Topological in-gap boundary states exist due to nontrivial topological invariants. Middle) Non-Hermitian skin states occur when non-reciprocal pumping forces \emph{all} states towards a boundary. %, and are epitomized by completely different open vs. periodic boundary condition (OBC vs. PBC) spectra. 
Right) By contrast, EB states require neither nontrivial topology nor the skin effect, existing due to the defectiveness of a non-Hermitian critical point. Their occupation probabilities $p$ are enigmatically either negative or greater than one, unlike other eigenstates %of the bounded propagator matrix 
of $\bar P$ whose eigenvalues $p$ are clustered around 0 or 1. Also unlike topological or skin states which exponentially decay from the boundary, EB states are distinctively shaped by the singularity structure at the exceptional point.
}
\label{fig:1}
\end{figure}

\noindent{\textit{Exceptional Boundary states from defectiveness. --}}
To understand why exceptional points give rise to EB states, we consider a simple paradigmatic 2-level 1D Hamiltonian containing an EP~\footnote{More generic EPs can be obtained from it via basis transformation~\cite{shen2018topological}.}, indexed by momentum $k$:
\begin{equation}
H(k)=\left(\begin{matrix}
\gamma(k) & a(k) \\
b(k) & -\gamma(k)
\end{matrix}\right)\xrightarrow[k\rightarrow 0]{}\left(\begin{matrix}
\gamma_0 k^\Gamma & a_0 \\
b_0 k^B & -\gamma_0 k^\Gamma
\end{matrix}\right)
\label{EP}
\end{equation}
\thickmuskip=4mu
whose eigenenergies are $\pm \epsilon(k)=\pm \sqrt{a(k)b(k)+\gamma^2(k)}$, centered around the Fermi level at $\epsilon(k)=0$.  At the EP $k=0$, we require $a(0)=a_0\neq 0$ but $b(0),\gamma(0)\rightarrow 0$. For concreteness, we stipulate that $b(k)\rightarrow b_0k^B$ and $\gamma(k)\rightarrow \gamma_0k^\Gamma$ as $k\rightarrow 0$, where $B,\Gamma>0$. In this way, $H(0)$ is defective, possessing only one right eigenvector $(1,0)^T$. Its complement $(0,1)^T$ is also needed to span the Hilbert space, but disappears from the eigenspace as the EP is approached. 

Physically, this EP defectiveness causes the fermionic propagator $\langle c^\dagger_{x_1,\alpha}c_{x_2,\beta}\rangle = \langle x_1,\alpha|P|x_2,\beta\rangle=\sum_k P^{\alpha\beta}(k) e^{ik(x_1-x_2)}$ to be singular. Here
\begin{equation}
P(k)=|\phi^R(k)\rangle\langle \phi^L(k)|=\frac1{2}\left(\mathbb{I}-\frac{H(k)}{\epsilon(k)}\right)
\label{Pk}
\end{equation}
is the biorthogonal projector that projects onto the occupied band $|\phi^R(k)\rangle$ defined by $H|\phi^R\rangle = -\epsilon |\phi^R\rangle$, $H^\dagger|\phi^L\rangle = -\epsilon^* |\phi^L\rangle$, $\alpha,\beta$ labeling the sublattice indices. Ordinarily, away from an EP, $P(k)$ projects onto the well-defined occupied band. But at an EP, the occupied/unoccupied bands may not be resolvable in the defective eigenspace of $P$. Two possible classes of singular behavior exists, with the energy gap vanishing either like $\epsilon(k)\rightarrow \sqrt{a_0b_0}\,k^{B/2}$ or $\epsilon(k)\rightarrow  \gamma_0k^\Gamma$ depending on whether $B<2\Gamma$ or $B>2\Gamma$ ($\gamma_0=0$ is equivalent to setting $\Gamma\rightarrow \infty$ in this context). 

We shall first investigate the case of $0<B<2\Gamma$, where coupling asymmetry (in this basis choice~\footnote{A change in basis, say $\sigma_y\rightarrow \sigma_z$, will map coupling asymmetry to gain/loss. Note that this alone does not cause the non-Hermitian skin effect.}) dominates at small $k$, giving the simplest possible EP representation. In a translation invariant setting, this gives
\begin{equation}
P(k)\rightarrow %\frac1{2}\left(\begin{matrix}
%1 & -\sqrt{\frac{a_0}{b_0}}k^{-B/2} \\
%-\sqrt{\frac{b_0}{a_0}}k^{B/2} & 1
%\end{matrix}\right)=
\frac1{2}\left(\begin{matrix}
1 & -U(k) \\
-D(k) & 1
\end{matrix}\right)
\label{Pk2}
\end{equation}
where $U(k) =(D(k))^{-1} = \sqrt{\frac{a_0}{b_0}}k^{-B/2}$. Evidently, its eigenvalues are either $0$ or $1$ for all $k$, except at the EP $k=0$ where it is singular. In a finite system with $L$ sites, the smallest lattice momentum point is $k_0\sim \pi/L$, and the matrix element $P^{+-}(k_0)=-U(k_0)/2$ diverges like $\sim -\frac1{2}\sqrt{\frac{a_0}{b_0}}\left(\frac{L}{\pi}\right)^{B/2}$. The other matrix element $P^{-+}(k_0)=-D(k_0)/2$ vanishes correspondingly fast. Note that such divergences cannot exist at a Hermitian critical point, since the off-diagonal matrix elements would have to possess equal magnitudes, being complex conjugates. 

%can cause dramatic boundary accumulations that manifest when real-space inhomogeneity is introduced, i.e. through a potential well quench as elaborated later. 
EB eigenstates arise when this asymmetrically singular $P(k)$ acts in a spatially inhomogeneous setting i.e. a potential well quench or an entanglement subsystem. To understand the underlying mechanism, we first examine the divergence and asymmetry of the 2-site propagators by Fourier transforming $U(k)$ and $D(k)$ to real space, via $\langle c^\dagger_{x_1,+}c_{x_2,-}\rangle=-\frac1{2}\sum_k e^{ik(x_1-x_2)}U(k)$ and $\langle c^\dagger_{x_1,-}c_{x_2,+}\rangle=-\frac1{2}\sum_k e^{ik(x_1-x_2)}D(k)$. For concreteness, we introduce the ansatz
\begin{subequations}
\begin{eqnarray}
b(k)&=&b_0(2(1-\cos k))^{B/2},\\
a(k)&=&b(-k)+a_0
\end{eqnarray}
\end{subequations}
realizable with lattice hoppings across at most $B/2$ sites. Implementations for odd $B$ are discussed in ~\cite{SuppMat}. Near the EP, $\langle c^\dagger_{x_1,-}c_{x_2,+}\rangle \sim \sqrt{\frac{b_0}{a_0}}2^{B/2}e^{-4(\Delta x)^2/B}$ is short-ranged, quadratically decaying with $\Delta x=x_1-x_2$. However~\cite{SuppMat}, due to the divergent denominator in $U(k)\sim \sqrt{a_0/b(k)}$, 
\begin{equation}
\langle c^\dagger_{x_1,+}c_{x_2,-}\rangle|_{B> 4}\sim -\sqrt{\frac{a_0}{b_0}}\left(\frac{L}{\pi}\right)^{B/2-1}\times \left(2-\frac{\pi^2\,\Delta x^2}{L^2}\right),
\label{Bg4}
\end{equation}
which is long-ranged in $\Delta x = x_1-x_2$ and diverges as $L^{B/2-1}$. For the important cases of $B=2$ and $B=4$, we also have~\cite{SuppMat}
\begin{equation}
\langle c^\dagger_{x_1,+}c_{x_2,-}\rangle|_{B=2}\sim -\sqrt{\frac{a_0}{b_0}}\left(\log\frac{L}{\pi \Delta x}\right),
\end{equation}
\begin{equation}
\langle c^\dagger_{x_1,+}c_{x_2,-}\rangle|_{B= 4}\sim -\sqrt{\frac{a_0}{b_0}}\left(L-2\Delta x\right),
\label{B4}
\end{equation}
which diverges logarithmically and linearly with both $L$ and $x$. Due to the defectiveness of at the EP, even a very small asymmetry ($0\neq a_0\ll b_0$) at criticality can cause extremely large asymmetry in the hopping probability with sufficiently large $L$. 
%This is in spite of divergences 
%EB states occur when a boundary is abruptly imposed on a critical defective system with sufficiently divergent two-point function (???), i.e. $~k^{-B/2}$ where $B> ????????$. 

Evidently, then, a real-space cutoff will truncate $\langle c^\dagger_{x_1,+}c_{x_2,-}\rangle$ much more than $\langle c^\dagger_{x_1,-}c_{x_2,+}\rangle$, leading to uncompensated net accumulation. A distant caricature would be the
crossing of radiation across the extremal surface in the context of Hawking radiation~\cite{almheiri2020entropy,chen2020quantum}. Interestingly, this shall give rise to isolated EB states with special spatial profiles, alongside extensively many ordinary un-accumulated states (unlike the skin effect, which yields extensively many boundary accumulated states). 

\begin{figure}
\subfloat[]{\includegraphics[width=.94\linewidth]{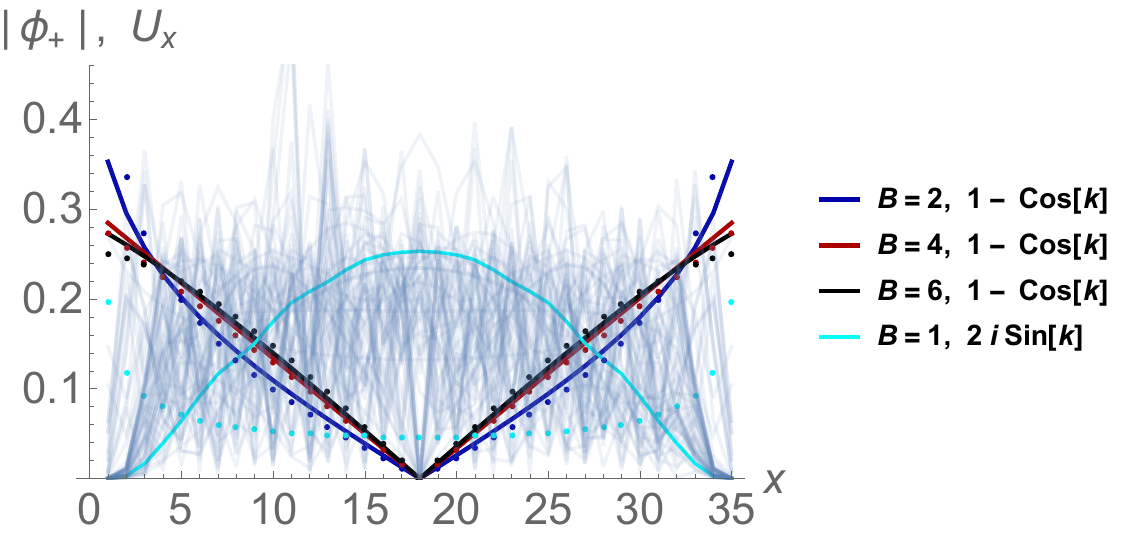}}\\
\subfloat[]{\includegraphics[width=.43\linewidth]{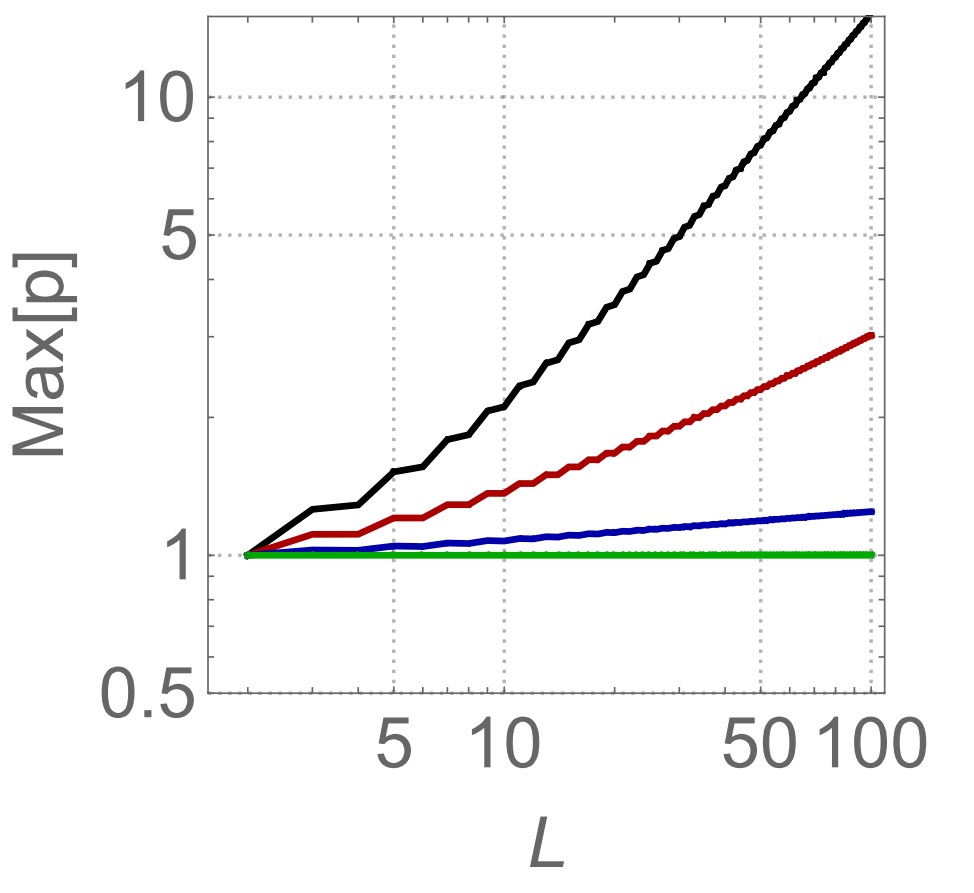}}
\subfloat[]{\includegraphics[width=.56\linewidth]{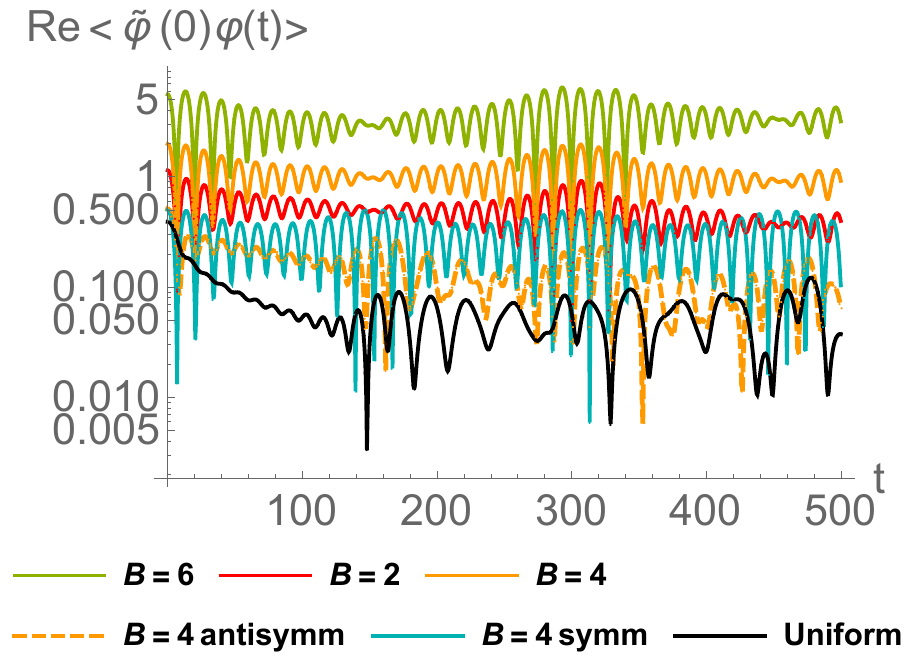}}
\caption{(a) Spatial profiles of EB states $\phi_+$ (solid) vs. the 2-site propagator $U_x=\langle c^\dagger_{x,+}c_{0,-}\rangle$. Close agreement is observed for sufficiently singular EPs with $b(k)\sim k^B$ for $B=2,4$ and $6$, but not the $B=1$ case given by $b(k)=2i\sin k$. Other non-EB states are superimposed as a gray background sea of states. (b) Anomalously large occupation probabilities $p=\left(1+\sqrt{1+\lambda}\right)/2$ of these EB states, colored as in (a). From Eq.~\ref{lambda}, they scale like $\sim L^{(B-4)/4}$ for sufficiently large $B$, but much more slowly for smaller $B$. In the $B=1$ case, which can be made Hermitian by setting $a_0=0$, $p\lessapprox 1$ just like ordinary projector eigenvalues. (c) Evolution of the 2-point function $\langle \tilde\varphi(0)\varphi(t)\rangle$ of various states $|\varphi\rangle$ with respect to a potential well quench with $\alpha=5$, time measured in units of $\omega_F^{-1}$. It remains elevated with fluctuations above unity when $|\varphi\rangle$ takes on the $B=2,4$ or $6$ EB states, scaling exponentially with $B$. However, it decays to significantly smaller values for non-EB states.}
\label{fig:2}
\end{figure}

Consider cutoffs at $x=0,\,l$ implemented by an operator $R$, which projects onto the region $0< x< l$, $l<L$. %~\footnote{It results in real-space cutoffs at $x=0$ and $x=l$, although their precise locations are unimportant.}. 
%To quantify how this cutoff at $x=0,l$ interplays with singular asymmetry, 
To quantify their effect, we introduce $\Lambda=4(\bar P^2-\bar P)=(\mathbb{I}-2\bar P)^2-\mathbb{I}$, where $\bar P=RPR$ is the $R$-truncated occupied band projector. % that has been truncated by $R$. While $\Lambda=P^2-P=0$ before the truncation, $\Lambda\neq 0$ after the truncation since $\bar P$ is no longer a projector. %$\Lambda$ measures how much $\bar P$ fails to be a projector due to the truncation (before truncation, $P^2=P$). 
Taking advantage of the purely off-diagonal form of $\mathbb{I}-2\bar P$ [Eq.~\ref{Pk2}], we simplify $\Lambda$ by decoupling it into $\Lambda = \Lambda_{+} \oplus \Lambda_{-}$, where $\Lambda_+=\bar U\bar D -\mathbb{I}$ and $\Lambda_-=\bar D\bar U-\mathbb{I}$. Due to the real-space cutoffs, $\bar U$ and $\bar D$ are $L\times L$ Toeplitz matrices that are not diagonal in $k$ and, importantly, no longer inverses of each other. In particular, since $\bar U$ contains long-ranged couplings while $\bar D$ contains only short-ranged couplings, there must be uncompensated off-diagonal elements in $\bar U\bar D$ near the cutoffs. As rigorously derived in the Supplement~\cite{SuppMat}, $\Lambda_+$ is mostly contributed by its boundary row and columns i.e. 
\begin{equation}
\Lambda_+\approx |c_1\rangle\langle 1|+|1\rangle\langle r_1|-\langle 1|c_1\rangle|1\rangle\langle 1|,
\label{lplus}
\end{equation}
where $|1\rangle $ is the site next to the $x=0$ cutoff and $|c_1\rangle$ and $\langle r_1|$ are the boundary row and column containing $|1\rangle$. %the first row and column of $\Lambda_+$ next to the cutoff. 
From $\Lambda_+=\bar U\bar D -\mathbb{I}$, their %boundary row and column 
elements are explicitly given by $\langle x|c_1\rangle = 4\sum_{j=0}\langle c^\dagger_{0,-}c_{1+j,+}\rangle\langle c^\dagger_{x+j,+}c_{0,-}\rangle$ which are slowly decaying, and $\langle r_1|x\rangle = 4\sum_{j=0}\langle c^\dagger_{0,+}c_{x+j,-}\rangle\langle c^\dagger_{1+j,-}c_{0,+}\rangle$ which are rapidly decaying (short-ranged). Importantly, this asymmetry %between the boundary row and column 
implies that it is the $|c_1\rangle\langle 1|$ term that dominates $\Lambda_+$, which must thus host a special eigenstate $|\phi_+\rangle \approx |c_1\rangle$. Left-multiplying Eq.~\ref{lplus} by $\langle 1 |$, we obtain $\Lambda_+|\phi_+\rangle \approx \lambda|\phi_+\rangle$ with eigenvalue $\lambda$ given by~\cite{SuppMat}
\begin{equation}
\lambda\approx \frac{\langle r_1|c_1\rangle}{\langle 1|c_1\rangle}=\frac{\langle 1|\Lambda_+^2|1\rangle}{\langle 1|\Lambda_+|1\rangle}\,\sim\, 
\begin{cases}
      L^{B/2-1}, & \quad B>2\\
      \log L, & \quad B=2\\
			\text{const}. & \quad B<2
    \end{cases}
		\label{lambda}
\end{equation}
Combined with $|\phi_-\rangle$, the analogous eigenstate of $\Lambda_-$ with the same eigenvalue, we obtain the Exceptional Boundary eigenstate $|\phi\rangle=|\phi_+\rangle\oplus|\phi_-\rangle$ of $\Lambda$. Since an eigenstate of $\Lambda=4(\bar P^2-\bar P)$ is also an eigenstate of $\bar P$, $|\phi\rangle$ is an EB eigenstate of the truncated projector $\bar P$ with eigenvalue $p=\left(1+ \sqrt{1+\lambda}\right)/2$. A corresponding EB eigenstate $|\phi'\rangle$ exists at the other cutoff boundary with eigenvalue $p'=1-p=\left(1-\sqrt{1+\lambda}\right)/2$. Physically, $p$ (or $p'$) is the occupation probability $\langle n\rangle=\langle\Phi^\dagger \tilde \Phi\rangle $ of the EB state $|\phi\rangle=\Phi^\dagger|0\rangle$ or its biorthogonal conjugate $|\tilde \phi\rangle=\tilde\Phi^\dagger|0\rangle$~\cite{SuppMat}:
\begin{eqnarray}
\langle n\rangle&=&  \sum_k\langle \psi^L(k)|\Phi^\dagger \tilde\Phi|\psi^R(k)\rangle=  \sum_k \langle \tilde\phi|\psi^R(k)\rangle\langle\psi^L(k)|\phi\rangle\notag\\
&=&  \langle \tilde\phi|P|\phi\rangle%=\langle \phi|RPR|\phi\rangle
=\langle \tilde\phi|\bar P|\phi\rangle=p.\qquad
\label{np}
\end{eqnarray}

\noindent{\textit{Unique properties of EB states. --}}
EB states are unusual because: (i) They are isolated eigenstates of $\bar P$ with occupation probabilities that are either negative ($p'<0$) or potentially much greater than one ($p>1$); (ii) Unlike conventional isolated mid-gap eigenstates which universally decay exponentially%away from the boundary
~\footnote{The exponential spatial decay length is universally given by the inverse imaginary gap~\cite{he2001exponential,lee2016band,lee2017band}.}, their spatial profiles resemble that of $\langle c^\dagger_{x_1,+}c_{x_2,-}\rangle$ [Eqs.~\ref{Bg4} to \ref{B4}], which can be quadratic, logarithmic or linear. This resemblance, which originates from $\langle x|\phi_+\rangle\approx \langle x|c_1\rangle\approx 4\langle c^\dagger_{0,-}c_{1,+}\rangle\langle c^\dagger_{x,+}c_{0,-}\rangle\propto \langle c^\dagger_{x,+}c_{0,-}\rangle$, is evident in Fig.~\ref{fig:2}a for sufficiently singular momentum profiles $b(k)\sim k^{B\geq 2}$.% with $B\geq 2$. %but not that of $B=1$ where the agreement is poor.

Anomalously large occupation probabilities $p>1$ of the EB states physically originate from the asymmetric accumulation caused by $ \langle c^\dagger_{x_1,-}c_{x_2,+}\rangle\neq \langle c^\dagger_{x_1,+}c_{x_2,-}\rangle$, which also diverge strongly with $\sim L^{(B-4)/4}$ for sufficiently large $B$ [Fig.~\ref{fig:2}b]. The corresponding negative occupation $p'=1-p$ of the EB states on the other boundary, away from the direction of asymmetry, represents a loss that compensates the large gain in particle density across the region. As previously explained, such spectacular non-locality cannot be brought about by non-defective critical points, which only lead to power-law decaying 2-site propagators and eigenstates, not the characteristic EB state behavior in Fig.~\ref{fig:2}a. This asymmetric inter-sublattice accumulation is also fundamentally distinct from the non-Hermitian skin effect~\cite{yao2018edge,xiong2018does,Lee2019anatomy,kunst2018biorthogonal}, which affects \emph{all} eigenstates, not just special EB states that are eigensolutions to Eq.~\ref{lplus}. Furthermore, a Pauli basis rotation, say, from $\sigma_y$ to $\sigma_z$ may transform this asymmetry into physical gain-loss without reciprocity breaking. Although classified by a number $B$, EB states are distinct from topological boundary states, whose robustness are rooted in topological anomalies rather than \emph{geometric} EP defectiveness. Besides, topological mid-gap states typically straddle the boundary with occupation $p\approx 1/2$~\cite{fidkowski2010entanglement,turner2010entanglement,alexandradinata2011trace,qi2012general,lee2015free}, unlike EB states whose $p,p'\notin[0,1]$.

\noindent{\textit{Persistent elevated EB 2-point functions. --}}
The EB state phenomenon leads to the existence of long-lived, high particle density configurations even after a critical system EP system is quenched. Typically, a post-quench Hamiltonian $H_F$ scrambles the eigenstates of the original Hamiltonian, eroding pre-quench initial states rapidly~\cite{sun2018uncover,yang2018dynamical,zhang2018dynamical,zhang2019characterizing,lee2020quenched,bacsi2020dynamics}. However, due to the very high $\langle n\rangle =p$ of EB states in large systems, the 2-point function $\langle \tilde\varphi(0)\varphi^\dagger(t)\rangle$ of EB states $|\varphi(0)\rangle$ %that approximately resemble EB states can still 
can persistent at elevated values. 

We consider a quench where the system evolves according to the EP Hamiltonian (Eq.~\ref{EP}) for times $t<0$, and a post-quench potential landscape $H_F=\sum_x E(x)|x\rangle\langle x|$ for $t\geq 0$. To demonstrate the significance of the bounded nature of EB states, we choose the spatial energy profile to be a potential well $E(x)=\hbar\omega_F/(1+e^{\alpha(l-x)})$ with a wall of slope $\alpha$ at $x=l$, the EB state cutoff. $\omega_F^{-1}$ sets the decay timescale expected of ordinary, non-EB states. 
%well Hamiltonian \red{change} $H_F=\hbar\omega_F\theta(x-l)\mathbb{I}=\hbar\omega_F(\mathbb{I}-R)$ for $t\geq 0$, where $\omega_F^{-1}$ sets the decay timescale for ordinary states. 
From $t=0$ onwards, the 2-point correlation function $\langle \tilde\varphi(0)\varphi(t)\rangle=\langle \varphi|e^{iH_Ft/\hbar}P|\tilde\varphi\rangle$ evolves from its initial value of $\langle n\rangle$, as plotted in Fig.~\ref{fig:2}c for various choices of $|\varphi\rangle=|\varphi(0)\rangle$ for the same well. Evidently, it remains elevated when $|\varphi\rangle$ is an EB state $|\phi\rangle$, scaling exponentially with $B$. In particular, for sufficiently singular EPs with $B>6$, the 2-point function shall remain robustly large even if the quench evolution maximally scrambles the states, since the dominant EB states take up a fraction of $2/L$ of all the states, but scales like $\langle n\rangle= (1\pm \sqrt{1+\lambda})/2\sim L^{(B-2)/4}>L$. However, the 2-point function of generic non-EB states $|\varphi\rangle$ remain in the range of $[0,1]$, as typical of ordinary expectation values in the absence of significant asymmetric gain/loss. This is the case even when $|\varphi(x)\rangle$ superficially resembles an EB state but is in reality almost orthogonal i.e. the symmetric $|\phi_+\rangle\oplus|\phi_+\rangle$ or antisymmetric $|\phi_+\rangle[\oplus|\phi_+\rangle]$ combinations, which are distinct from the actual EB state $|phi\rangle=|\phi_+\rangle\oplus|\phi_-\rangle$. The 2-point function decays even more substantially before fluctuating around the order of $1/\sqrt{L}$ when $|\varphi\rangle$ is completely different from any EB state i.e. when it is spatially uniform or random (see~\cite{SuppMat}).

%Note EB states are robust even \emph{without} the potential well, since Eq.~\ref{np} holds even in the absence of any boundary as long as $|\phi\rangle$ is compactly supported. % in $x\in [1,L]$. 

% ...Since an eigenstate $|\phi\rangle$ of $\bar P$ is also an eigenstate of $\bar P^2$ and hence $\Lambda$, writing $|\phi\rangle=|\phi_1\rangle\oplus|\phi_2\rangle$, we have $\Lambda_{jj}|\phi_j\rangle=\lambda|\phi_j\rangle$, $j=1,2$, with $\lambda$ their common eigenvalue. 

%Most prominently, defectiveness causes the divergence of the two-point function, which we now notate as $\langle c^\dagger_{x_1,\mu}c_{x_2,\nu}\rangle=\langle x_1,\mu|P|x_2,\nu\rangle$, $\mu,\nu=\pm$ labeling the sublattices. At the EP, $P(k)$ is singular, and these Fourier coefficients must diverge. Since the smallest lattice momentum point is $k_0\sim \pi/L$, we expect the inter-sublattice two-point function $\langle x_1,+|P|x_2,-\rangle$ to diverge like $\sqrt{\frac{a_0}{b_0}}\left(\frac{L}{\pi}\right)^{B/2}$. 

\noindent{\textit{Negative entanglement entropy from EB states. --}}
EB states also fundamentally modify free-fermion entanglement entropy and spectra. Consider a critical half-filled 1D Fermion gas (restricting ourselves to systems with real spectra) governed by the EP Hamiltonian of Eq.~1. A density operator consistent with probabilistic interpretations~\cite{brody2013biorthogonal,herviou2019entanglement,chang2020entanglement,SuppMat} is given by $\rho=|\Psi\rangle\langle \Psi|$, where $|\Psi\rangle$ is the many-body state and $\langle \Psi|$ its biorthogonal conjugate. To probe entanglement properties, we examine the reduced density matrix $\rho_{\scaleto{\mathcal{R}}{4pt}}=\text{Tr}_{\scaleto{\mathcal{R}^c}{4pt}}|\Psi\rangle\langle \Psi|$ obtained by tracing over the degrees of freedom outside an arbitrarily selected region $\mathcal{R}$, which we take as $0<x<l$ as before. From Wick's theorem~\cite{peschel2003calculation}, the EE $S=-\text{Tr}\rho_{\scaleto{\mathcal{R}}{4pt}}\log\rho_{\scaleto{\mathcal{R}}{4pt}}$ and the entanglement Hamiltonian $H_E=-\log \rho_{\scaleto{\mathcal{R}}{4pt}}$ can be expressed in terms of operators in the single-particle Hilbert space: $S=-\text{Tr}[\bar P\log\bar P +(\mathbb{I}-\bar P)\log(\mathbb{I}-\bar P)]$ and $H_E=\log[(\mathbb{I}-\bar P)/\bar P]$, where $\bar P=RPR$ is the $R$-truncated occupied band projector introduced earlier. Hence an EB eigenstate satisfying $\bar P |\phi\rangle=p|\phi\rangle$ must provide an additional $\epsilon_\phi=\log[p^{-1}-1]$ contribution to the entanglement spectrum, which will be complex since $\text{Re}\,p>1$. The corresponding EB eigenstate on the opposite boundary of $\mathcal{R}$, with eigenvalue $p'=1-p$, gives rise to an opposite entanglement eigenenergy $-\epsilon_\phi$. 

\begin{figure}
\includegraphics[width=\linewidth]{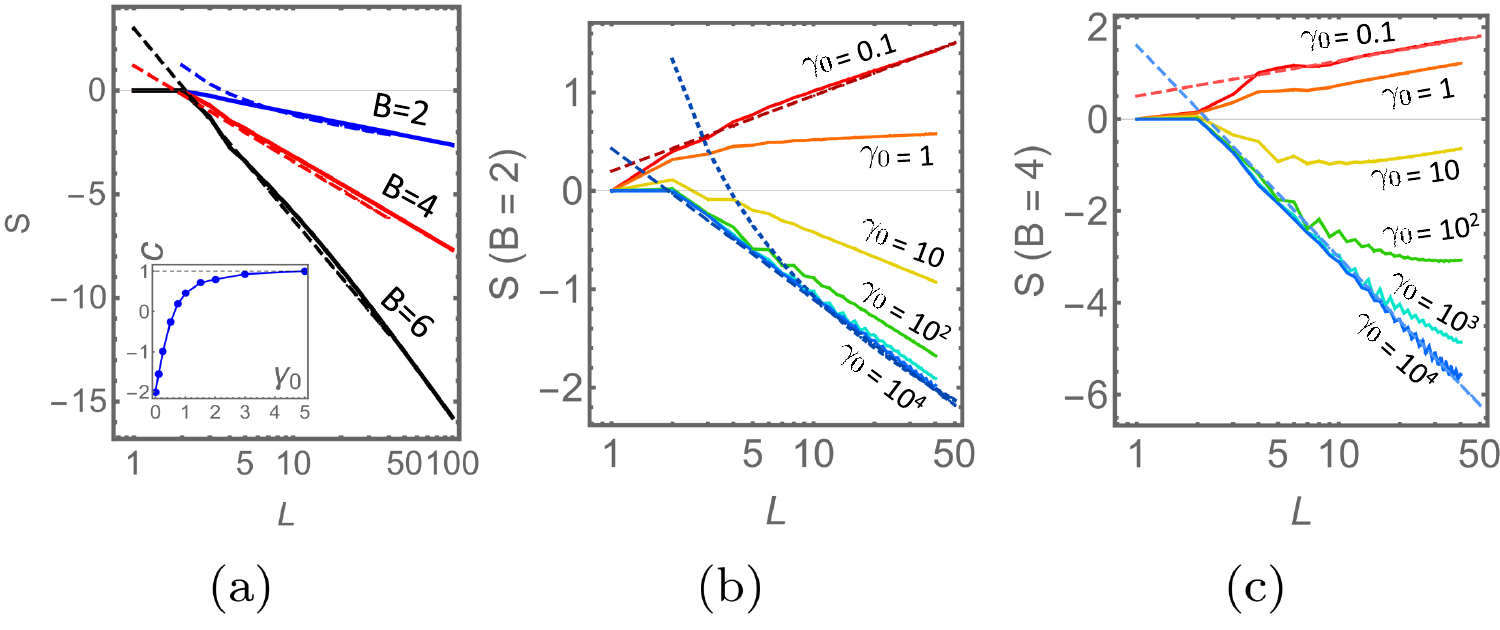}
%\subfloat[]{\includegraphics[width=.32\linewidth]{fig3a.pdf}}
%\subfloat[]{\includegraphics[width=.33\linewidth]{fig_gamma_B=2.pdf}}
%$\,$
%\subfloat[]{\includegraphics[width=.33\linewidth]{fig_gamma_B=4.pdf}}
%\subfloat[]{\includegraphics[width=.17\linewidth]{c_gamma0.pdf}}
\caption{(a) Negative logarithmic scaling of the EE $S$ due to ideal ($\gamma_0=0$) EB states. Numerical results (solid) agree well with predicted asymptotic behavior $S=\frac{c}{3}\log L$ (dashed), with $c=3(2-B)$ for $B>2$ [Eq.~\ref{S}] and $c=-2$ for $B=2$. Inset: Variation of $c$ with the tuning of $\gamma_0$ for $H'$ [Eq.~\ref{Hvwg}] at $w=2,v=2.5$. (b) Continuous tuning of the logarithmic scaling coefficient of $S$ as $\gamma_0$ is varied at large $a_0=10^3$ and $b_0=1$. Dashed straight lines represent $S\sim\log L$ (ordinary critical) and $S\sim-\frac{2}{3}\log L$ (EB) fits, while the dashed curve represent the qualitative fit with $S_\text{EB}\approx 0.6-2\log\log L$ as derived in~\cite{SuppMat}. (c) Scaling behavior of $S$ for $B=4$, also at $a_0=10^3,b_0=1$. Unlike in $B=2$, $c$ does not change continuously as $\gamma_0$ is varied, but crossovers from $\sim\log L$ (red dashed) to the EB behavior of $\sim -2\log L$ (blue dashed) at a $\gamma_0$-dependent critical size.
}
\label{fig:3}
\end{figure}

Most interestingly, EB eigenstates lead to negative contributions to the EE $S_\text{EB}=-p\log p -(1-p)\log(1-p)$. Even though $\log p$ and $\log(1-p)$ are complex for $\text{Re}\,p>1$, they always combine to produce a real and negative $S_\text{EB}$. Since $\lambda \sim L^{B/2-1}$ in $p=(1+\sqrt{1+\lambda})/2$ for $B>2$, 
\begin{equation}
S_\text{EB}|_{B>2} \sim \log 4 -2 -\log \lambda_0-\frac{B-2}{2}\log L,
\label{S}
\end{equation}
the constant $\lambda_0$ computable in terms of the 2-site propagators~\cite{SuppMat}. In the absence of topological contributions, most of the other eigenvalues of $\bar P$ are extremely close to $0$ or $1$, and contribute negligibly to $S$. Hence $S\approx S_\text{EB}$ for EP Hamiltonians [Eq.~1], with its negative $\log L$ dependence accurately corroborated by numerical computations across various $B$ [Fig.~\ref{fig:3}a]~\footnote{The $B=2$ case has slow convergence subtleties~\cite{SuppMat}, and we numerically obtain $S\sim-2/3\log L$.}. Physically, the negative EE originates from the anomalously large expected EB occupancy $\langle n\rangle =p>1$ (and $1-\langle n\rangle =p'<0$ for the opposite EB state), which is encoded in the reduced density matrix as: 
\begin{equation}
\rho_{\scaleto{\mathcal{R}}{4pt}}=\left(p|\phi\rangle\langle \tilde\phi|+p'|0\rangle\langle 0|\right)\otimes\left(p'|\phi'\rangle\langle \tilde\phi'|+p|0\rangle\langle 0|\right)\otimes \rho_{\scaleto{\mathcal{R}}{4pt}}^\text{EB'}
%\bigotimes_{\xi\neq p} \left(r_\xi|\chi'_{\scaleto{\xi}{4pt}}\rangle\langle\chi'_{\scaleto{\xi}{4pt}}|+(1-r_\xi)|0\rangle\langle 0|\right).
\end{equation}
where $\rho_{\scaleto{\mathcal{R}}{4pt}}^\text{EB'}=\bigotimes_{j\notin \text{EB}}\left(p_j|\psi_j\rangle\langle\tilde\psi_j|+(1-p_j)|0\rangle\langle 0|\right)$ is spanned by the \emph{non}-EB states $|\psi_j\rangle$ corresponding to all other $\bar P$ eigenvalues $p_j\in[0,1]$. In particular, in the large $B$ or $L\rightarrow \infty$ thermodynamic limit where $p\gg 1$, we have $\rho_{\scaleto{\mathcal{R}}{4pt}}\approx -p^2\sigma_z^{\phi}\otimes \sigma_z^{\phi'}\otimes\rho_{\scaleto{\mathcal{R}}{4pt}}^\text{EB'}$, where $\sigma_z^\phi=|\phi\rangle\langle\tilde \phi|-|0\rangle\langle 0|$. Intuitively, the coefficient of $|0\rangle\langle 0|$ indicates how much the state is entangled with degrees of freedom outside of $\mathcal{R}$. By contrast, this coefficient is very close to $0$ or $1$ for gapped systems with short-ranged entanglement, since the state on different sides of the entanglement cut can be almost completely decoupled. For usual gapless systems, there also exists some coefficients nearer to $1/2$ due to long-ranged entanglement. But for EB states, we not just have long-ranged entanglement, but also \emph{negative} effective entanglement due to EB states having $\bar P$ eigenvalues $p,p'\notin[0,1]$. This unusual scenario occurs due to the divergent unbalanced 2-site propagators, which can drastically amplify or attenuate a state across an entanglement cut.

%The boundary of $\mathcal{R}$ is known as the entanglement cut that truncate the region outside of $0<x<l$ can be implemented by the $R$ projector introduced earlier.

\noindent{\textit{Generalizations and tunable entanglement scaling. --}}
So far, we have only  studied ``ideal'' EPs with the minimal ingredient necessary for defectiveness i.e. $\sim k^B$ dispersion with $\sim a_0$ asymmetry. But actual non-Hermitian critical systems often contain additional complications, such as nontrivial $\gamma(k)$ [Eq.~1] that breaks sublattice symmetry (or other symmetries upon basis rotation). Since $\gamma(k)$ enter the $\epsilon(k)$ denominator in $P(k)$ [Eq.~2], $U(k)\neq (D(k))^{-1}$ and the 2-site propagators $\langle c^\dagger_{x_1,\pm}c_{x_2,\mp}\rangle$ may no longer be distinctively long/short-ranged. This attenuates the first column of $\Lambda_+$ [Eq.~\ref{lplus}], leading to the erosion of EB states. 

For illustration, consider the following non-Hermitian SSH-like model generalized from Ref.~\cite{chang2020entanglement}: 
\begin{equation}
H'(k)=(v-w\cos k)\sigma_x + \gamma_0\sin k \,\sigma_y+i(v-w)\sigma_z.
\label{Hvwg}
\end{equation}
After swapping $\sigma_y$ and $\sigma_z$, it takes the form of Eq.~1 with $B=2,a_0=2(v-w),b_0=w/2$. It straddles the transition between the topologically trivial and gapless phase, and hosts an EP at $k=0$, albeit with a diagonal ($\sigma_z$) term $\gamma(k)=\gamma_0\sin k$. As $\gamma_0$ is increased from $0$ (with $v,w$ kept fixed), the coefficient $c$ in its EE $S\sim \frac{c}{3}\log L$ increases from $-2$ to $1$ [Fig.~\ref{fig:3}a inset]. The negative $c=-2$ coefficient is a consequence of the marginally divergent EB state, and has been previously explained in terms of non-unitary CFT for a related model~\cite{chang2020entanglement}. Increasing $\gamma_0$ erodes the EB state till we finally obtain $c=1$ as expected of an ordinary critical free fermion. In a ``stronger'' or more asymmetric $B=2$ EP with large $a_0$ [Fig.~\ref{fig:3}b], $c$ can be finely tuned by changing the magnitude of $\gamma_0$. For $B>2$, however, $S$ crossovers from $c=1$ to the asymptotic value of $c=-3(B-2)$ as $L$ increases, and tuning $\gamma_0$ can only changing the critical crossover system size $L$, not $c$ itself [Fig.~\ref{fig:3}c]. Various representative models that host EB states are tabulated in~\cite{SuppMat}.

More generally, systems with multiple EPs can possess multiple sets of EB states with their own elevated $\langle n\rangle$ and 2-point functions. These are distinct from \emph{secondary} EBs for a single EP of sufficiently large $B$, which exist due to increased hopping non-locality and possess related spatial profiles with $\lambda\sim L^{B/2-2}$~\cite{SuppMat}. Remnants of EB states can even exist for an EP with a small gap $m$, although $\lambda$ will scale like $m^{1-B/2}$ instead of $L^{B/2-1}$.

%\noindent{\textit{OBC cases, skin effect EPs.... --}}

%\red{- talk about OBC vs PBC cases, particularly for EE. Need to explore how the scaling behaves when there is the skin effect (is it trivial, or deserving of a separate paper?). The eigenstates obey the boundary conditions, regardless of whether $R$ has been applied. }\\

%\red{- what if topo modes and EB modes coincide? Usually not possible as one of them is gapped/gapless, but let's think more about it first. What about interplay between the skin effect and EP effects? talk about NHSE EPs}

\noindent{\textit{Discussion. --}}
EB states represent a new type of robust boundary phenomena distinct from topological or non-Hermitian skin states. They fundamentally arise because EP defectiveness causes the propagator to be asymmetrically singular, such that special bounded states can possess anomalously large occupation probabilities which may be detectable in optical lattice setups. Such EB states also %admit negative effective probabilities in the Schimdt decomposition of the reduced density matrix, leading to potentially negative EE.    
contribute negatively to the EE, with the logarithmic scaling coefficient $c$ depending on the exact EP dispersion. Whether $c$ can be identified with the central charge of a CFT remains an open question. 

%MENTION TABLE EARLY ON..A table of various EB state models and their parameters.
%Yet, instead of necessarily requiring \emph{physical} boundaries,% Finally,  and exist independently of the presence of physical boundaries. 
\noindent{\textit{Acknowledgements --}} We thank Xueda Wen, Jiangbin Gong and Zhengliang Lim for helpful discussions.

%\red{.tunable EE....whether $c$ is the central charge....open question...need to find CFT for that first..last section table}
\bibliography{references}

\clearpage

\onecolumngrid
\begin{center}
\textbf{\large Supplementary Materials}\end{center}
\setcounter{equation}{0}
\setcounter{figure}{0}
\renewcommand{\theequation}{S\arabic{equation}}
\renewcommand{\thefigure}{S\arabic{figure}}
\section{I. Fundamental properties of Exceptional Boundary (EB) eigenstates}

\subsection{Boundary-truncated Projector}

We first show why EB states emerge as the eigenstates of the boundary-restricted two-point function, which is defined as $\bar P=RPR$, where $P$ is the biorthogonal occupied band projector of the translation-invariant exceptional system and $R$ the projector onto the real space region demarcated by the boundary. For general particle-hole symmetric Hamiltonians of the form
\begin{equation}
H=\left(\begin{matrix}
0 & a \\
b & 0 
\end{matrix}\right)=\sum_k\left(\begin{matrix}
0 & a(k) \\
b(k) & 0 
\end{matrix}\right)|k\rangle\langle k|
\end{equation}
in two-component sublattice space, the biorthogonal projector onto the occupied band $\{|\psi^R(k)\rangle\}$ takes the form
\begin{equation}
P = \sum_k|\psi^R(k)\rangle\langle \psi^L(k)|=\frac1{2}\left(\begin{matrix}
\mathbb{I} & -\sqrt{\frac{a}{b}} \\
-\sqrt{\frac{b}{a}}  & \mathbb{I} 
\end{matrix}\right)
=\frac1{2}\left(\begin{matrix}
\mathbb{I} & -U \\
-D & \mathbb{I} 
\end{matrix}\right)
%P(k) = \frac1{2}\left(\begin{matrix}
%1 & -u(k) \\
%-d(k) & 1 
%\end{matrix}\right)
\end{equation}
where $\langle \psi^L(k)|$ is the biorthogonal dual of $|\psi^R(k)\rangle$, and $U\neq D^\dagger$ for a non-Hermitian system. In a translation invariant setting, we can work in momentum space where we have $U(k)D(k)=1$, and $P$ only has eigenvalues $0$ and $1$. This holds because the divergence in $U(k)\rightarrow \sqrt{\frac{a_0}{b_0}}k^{-B/2}$ is canceled by the vanishing of $D(k)\rightarrow \sqrt{\frac{b_0}{a_0}}k^{B/2}$ as we approach the EP at $k=0$ (Eq.~3 of the main text). 

But when projected onto a bounded region in real space i.e. $u\rightarrow \bar U = RUR$ and $d\rightarrow \bar  D = RDR$, we have $\bar U\bar  D \neq \bar  D\bar U \neq 1$. Here $R$ truncates off contributions to $\bar U$ or $\bar D$ from outside the stipulated real space region, such that $\bar U$ and $\bar D$ are no longer inverses of each other. The departure of $\bar U \bar D $ or $\bar D \bar U$ from unity captures the effect of boundaries or, more generally, spatial inhomogeneities. To study this systematically, we introduce the quantity $\Lambda=(\mathbb{I}-2\bar P)^2-\mathbb{I}=4(\bar P^2-\bar P)$, which also measures how much $\bar P\neq \bar P^2$ fails to be a projector. Since $\Sigma =R(PRP-P)R=-RPR^c PR$ where $ R^c =\mathbb{I}-R$ projects onto the complement of region $R$, $\Sigma$ can also be understood as a measure of how much the occupied band projector $P$ ``entangles'' $R$ and $R^C$ (If all band were occupied, $\Sigma$ would be equal to $RR^cR=0$). Its relation with EE will be elaborated later on.

 Due to the off-diagonal form of $\mathbb{I}-2\bar P$, $\Lambda$ is diagonal, such that its eigenstates $|\phi\rangle=(|\phi_+\rangle,|\phi_-\rangle)^T$ decouple into $\pm$ sublattice subspaces:
\begin{equation}
\Lambda\left(\begin{matrix}
|\phi_+\rangle \\
|\phi_-\rangle
\end{matrix}\right)
= \left(\begin{matrix} 
\Lambda_+ & 0 \\
0 & \Lambda_-
\end{matrix}\right)
\left(\begin{matrix}
|\phi_+\rangle \\
|\phi_-\rangle
\end{matrix}\right)
= \left(\begin{matrix}
\bar U\bar  D-\mathbb{I} & 0 \\
0 & \bar D\bar U-\mathbb{I}
\end{matrix}\right)
\left(\begin{matrix}
|\phi_+\rangle \\
|\phi_-\rangle
\end{matrix}\right)=\lambda\left(\begin{matrix}
|\phi_+\rangle \\
|\phi_-\rangle
\end{matrix}\right).
\label{Lambdaa}
\end{equation}
In particular, from the definition $\Lambda=4(\bar P^2-\bar P)$, each eigenvalue $\lambda$ of $\Lambda$ (or $\Lambda_\pm$) is related to an eigenvalue $p$ of the projector via $\lambda =4p(p-1)$, and can be obtained by working in either subspaces. These eigenvalues would have been trivially zero if not for the projection onto the bounded region, where $p$ cease to be $0$ or $1$.

Concretely, $\bar U$ and $\bar D$ can be obtained from $U$ and $D$ by a real-space truncation. We Fourier expand them into real space matrix elements via
\begin{align}
U(k)&=\sum_x U_x e^{-ikx}\\
D(k)&=\sum_x D_x e^{-ikx},
\end{align}
such that the two-point function takes the form
\begin{equation}
\langle x_1|P|x_2\rangle = \sum_k P(k) e^{ik(x_1-x_2)}=\frac1{2}\left(\begin{matrix}
\delta_{x_1,x_2} & -\sum_x U_x \delta_{x_1,x_2+x} \\
-\sum_x D_x \delta_{x_1,x_2+x} & \delta_{x_1,x_2} 
\end{matrix}\right)=\frac1{2}\left(\begin{matrix}
\delta_{x_1,x_2} & - U_{x_1-x_2} \\
-D_{x_1-x_2} & \delta_{x_1,x_2} 
\end{matrix}\right).
\end{equation}

 Since $\langle x_1|(\mathbb{I}-2P)|x_2\rangle=\left(\begin{matrix}
0 &  U_{x_1-x_2} \\
D_{x_1-x_2} & 0
\end{matrix}\right)$, when restricted to finite values $0\leq x_1,x_2 < L$ i.e. $P\rightarrow \bar P$, we have, for a single boundary at $x_1,x_2=0$,
\begin{equation}
\langle x_1|\Lambda|x_2\rangle=\langle x_1|(\mathbb{I}-2\bar P)^2-\mathbb{I}|x_2\rangle =\left(\begin{matrix}
\sum_{\substack{x_d\leq x_1,\\x_u\leq x_2}} U_{-x_u}D_{x_d}\delta_{x_u+x_d,x_1-x_2} -\delta_{x_1,x_2}& 0\\
0 & \sum_{\substack{x_u\leq x_1,\\x_d\leq x_2}} D_{x_d}U_{-x_u}\delta_{x_u+x_d,x_1-x_2}-\delta_{x_1,x_2}
\end{matrix}\right).
\label{Im2P2}
\end{equation}
This is the explicit form of $\Lambda$ in Eq.~\ref{Lambdaa} in terms of real space matrix elements. The above summations keep track of the terms that are truncated in the convolution of the Fourier series of $U(k)$ and $D(k)$. Had no boundaries been present, the convolutions would have evaluated to $\delta_{x_1-x_2}\mathbb{I}$, since $U(k)D(k)=D(k)U(k)=1$. In the presence of another boundary at $x_1,x_2=L$, we will also have analogous upper limits in the summations above.

\subsection{Two-site propagators in the singular limit}

If both $U(k)$ and $D(k)$ were smooth and non-divergent, their Fourier components (2-site propagators) would decay rapidly, rendering boundary effects largely negligible beyond small edge corrections. This scenario corresponds to $\langle x_1|\Lambda|x_2\rangle = \langle x_1|[(\mathbb{I}-2\bar P)^2-\mathbb{I}]|x_2\rangle\approx 0$. However, if $U(k)$ and $D(k)$ possess divergent behavior, they will affect the 2-site propagators defined as their Fourier transforms:
\begin{subequations}
\begin{align}
\langle c^\dagger_{x_1,+}c_{x_2,-}\rangle &= \langle x_1,+|P|x_2,-\rangle = -\frac1{2}U_{x_1-x_2}=-\frac1{2}\sum_k e^{ik(x_1-x_2)}U(k)\\
\langle c^\dagger_{x_1,-}c_{x_2,+}\rangle &= \langle x_1,-|P|x_2,+\rangle = -\frac1{2}D_{x_1-x_2}=-\frac1{2}\sum_k e^{ik(x_1-x_2)}D(k)\\
\langle c^\dagger_{x_1,\pm}c_{x_2,\pm}\rangle &=\frac1{2}.
\end{align}
\end{subequations}
Evidently, only the inter-sublattice 2-site propagators are nontrivial (this is due to the absence of any $\sigma_z$ term in $H$, whose complicating effects are treated in a later section.) As explained shortly, $U(k)$ diverges like $\sqrt{\frac{a_0}{b_0}}\left(\frac{L}{\pi}\right)^{B/2-1}$ for $B>2$ due to the EP defectiveness. For $B=2$, we shall also show below that $U(k)$ diverges like $\log L$. On the other hand, $D(k)$ remains well-behaved, and will for realistic local systems contain not more than a few non-neglible Fourier coefficients. 

\subsubsection{Ansatz for $a(k)$ and $b(k)$}

\begin{figure}
\subfloat[]{\includegraphics[width=.49\linewidth]{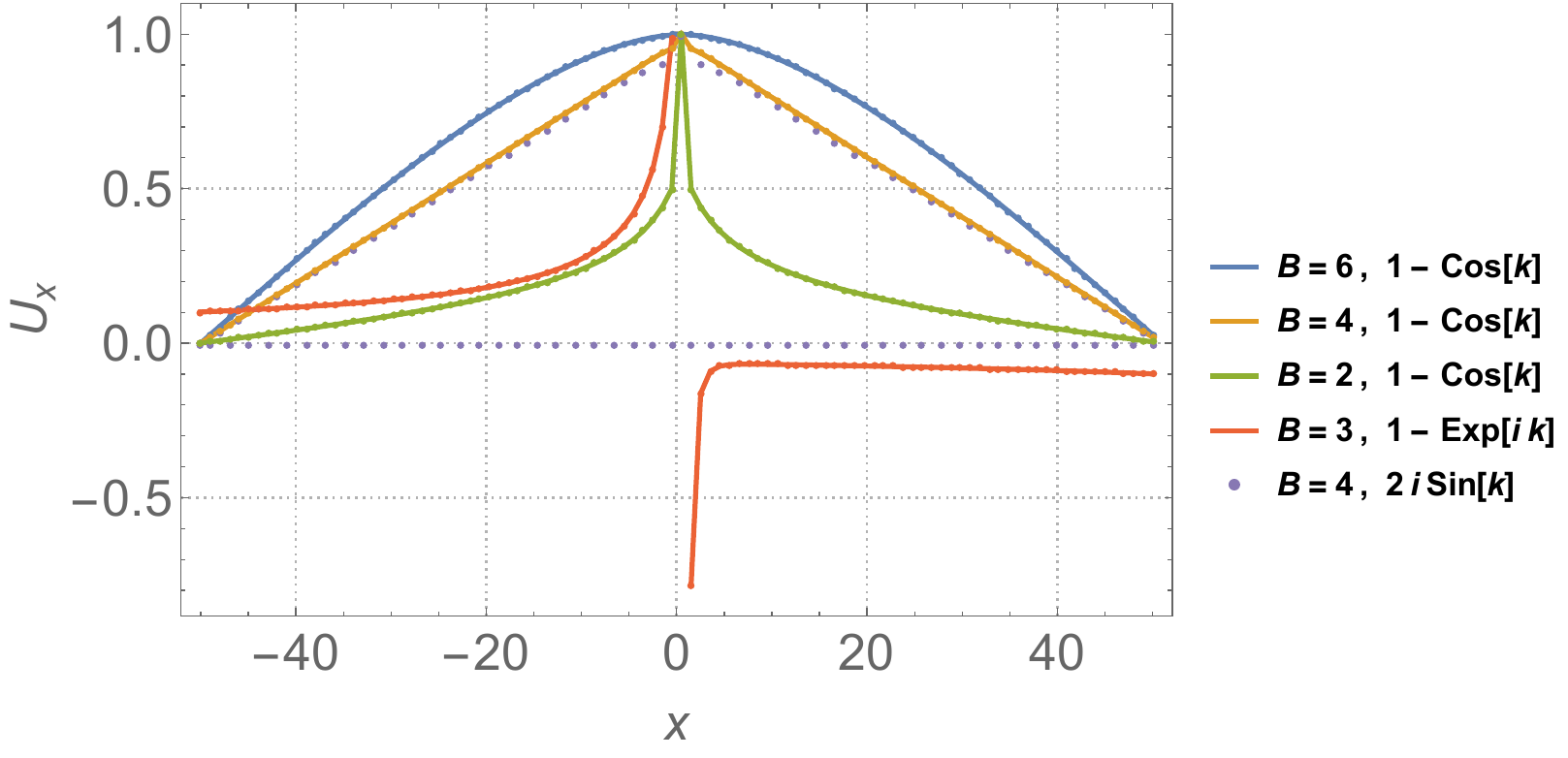}}
\subfloat[]{\includegraphics[width=.49\linewidth]{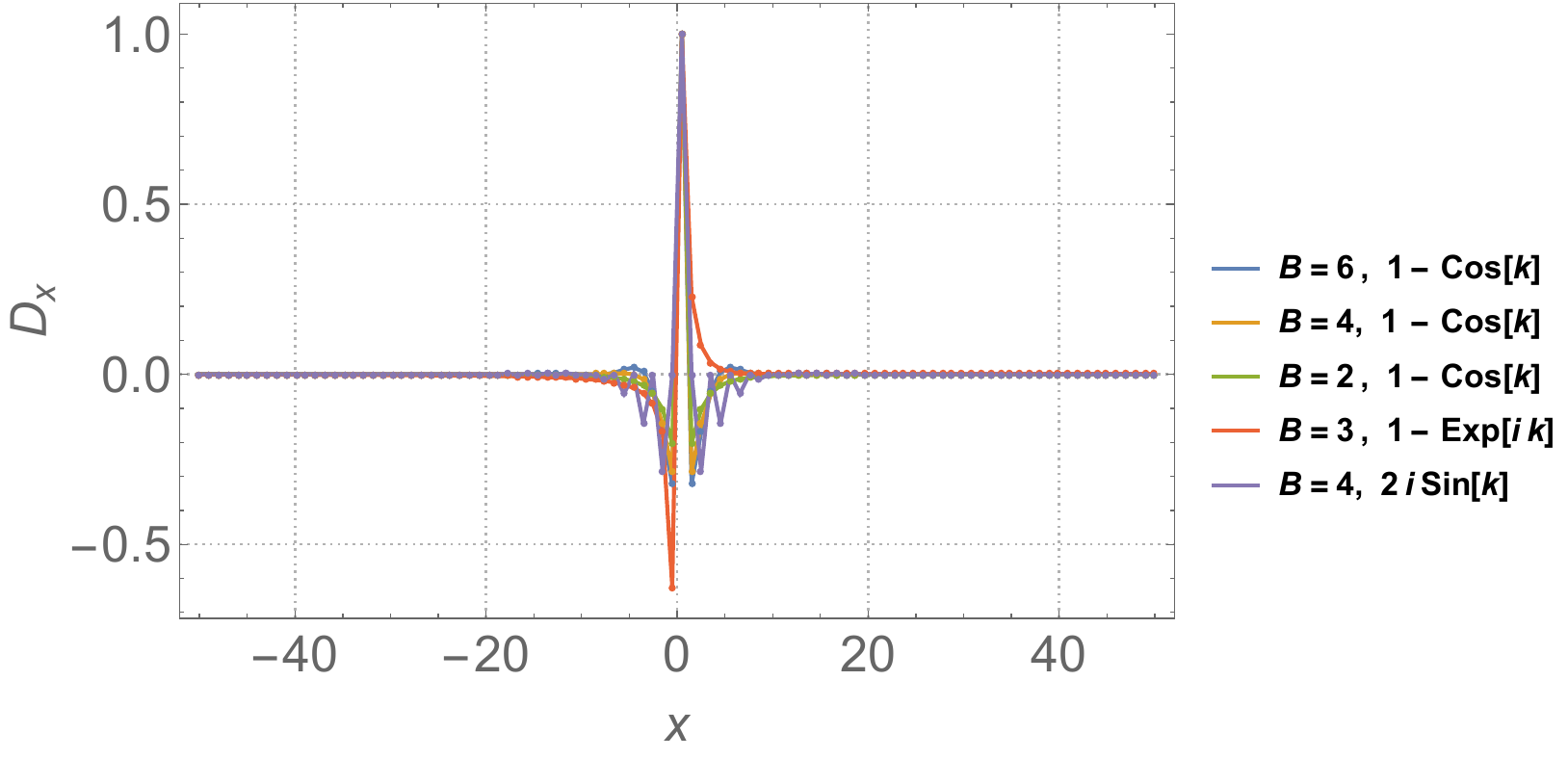}}\\
\subfloat[]{\includegraphics[width=.46\linewidth]{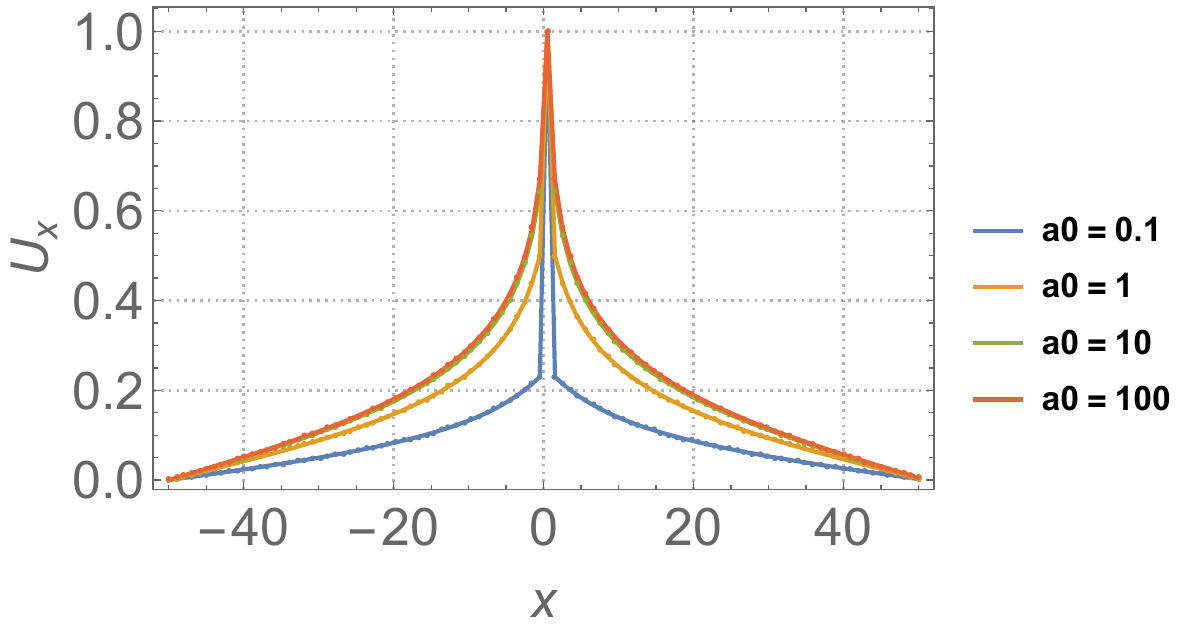}}
\subfloat[]{\includegraphics[width=.49\linewidth]{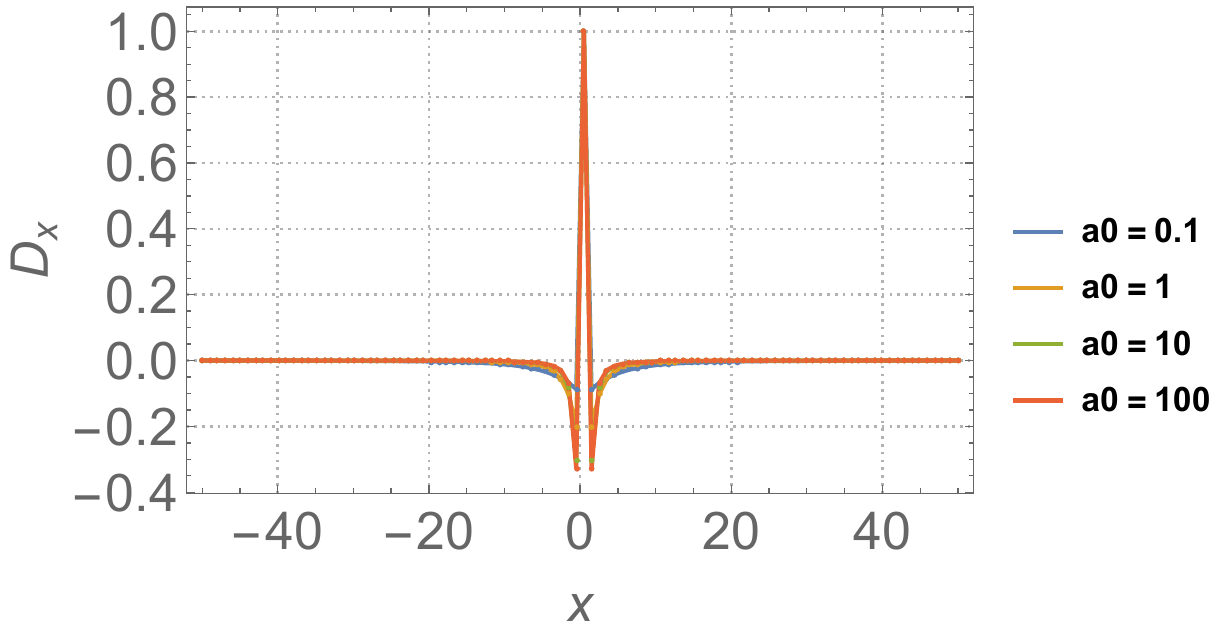}}
\caption{
Comparison between the two-point functions $\langle c^\dagger_{x,+}c_{0,-}\rangle = -U_x/2$ and $\langle c^\dagger_{x,-}c_{0,+}\rangle = -D_x/2$ of various ansatz exceptional points. (a) and (b) respectively depict $U_x$ and $D_x$ (both normalized to have a maximum value of $1$) for EPs defined by $b(k)=(1-e^{ik})^3$, $(2i\sin k)^4$ and $(2(1-\cos k))^B$, $B=2,4,6$, where $a(k)=b(-k)+a_0$ with $a_0$ set to unity. Due to EP defectiveness, $U_x$ is evidently more nonlocal than $D_x$, exhibiting power-law, linear and logarithmic decays ($B=3,6$, $B=4$ and $B=2$ respectively). They scale with $x$ in an exactly analogous fashion as with $L$ (Eqs.~\ref{B1} to \ref{B6}). (c-d) dependence of $U_x$ and $D_x$ on the EP asymmetry $a_0$. $D_x$, being already local, is practically independent of $a_0$. But $U_x$ depends significantly on $a_0$ for $a_0\sim 1$ or smaller, when our analytical results become no longer valid.
} 
\label{figa:fouriercompare}
\end{figure}

To be concrete, we specialize to the ansatz 
\begin{subequations}
\begin{align}
b(k) &=b_0(2(1-\cos k))^{B/2}\\
a(k)&= b(-k)+a_0,
\label{abk}
\end{align}
\end{subequations}
such that 
\begin{align}
U(k)&= \sqrt{\frac{a(k)}{b(k)}}=\sqrt{\frac{b_0(2(1-\cos k))^{B/2}+a_0}{b_0(2(1-\cos k))^{B/2}}}\\
D(k)&= \sqrt{\frac{b(k)}{a(k)}}=\sqrt{\frac{b_0(2(1-\cos k))^{B/2}}{b_0(2(1-\cos k))^{B/2}+a_0}}.
\end{align}
Of course, there are many other valid ansatze i.e. $b(k)=b_0\sin^Bk$ or $b_0(1-e^{ik})^B$, but Eq.~\ref{abk} possesses the most local Fourier coefficients, as shown in Fig.~\ref{figa:fouriercompare}. Near the EP at $k=0$, $a(k)$ is dominated by $a_0$, and $D(k)\approx \sqrt{\frac{b_0}{a_0}}(2(1-\cos k))^{B/4}=\sqrt{\frac{b_0}{a_0}}\sum_{\Delta x=-B/4}^{B/4}\binom{B/2}{B/4+\Delta x}(-1)^le^{i\Delta xk}$ i.e. with Fourier coefficients $(-1)^{\Delta x}\sqrt{\frac{b_0}{a_0}}\binom{B/2}{B/4+\Delta x}\sim (-1)^\Delta x\sqrt{\frac{b_0}{a_0}}2^{B/2}e^{-4(\Delta x)^2/B}$ that are rapidly decaying with $\Delta x$. Yet, because $U(k)\sim \sqrt{\frac{a_0}{b_0}}(2(1-\cos k))^{-B/4}$ is highly singular, its Fourier coefficients are slow-decaying and divergent with $L$. Since the closest momentum point to the EP is $k_0\sim \pi/L$, $a(k)$ is dominated by $a_0$ only when $(2(1-\cos k_0))^{B/2}=k_0^B\ll a_0$ i.e. $L\gg \pi/a_0^{1/B}$. In this limit, which is easily satisfied for modest system sizes of $L\sim 10^1-10^2$, we have 
\begin{eqnarray}
U_x=-2\langle c^\dagger_{x,+}c_{0,-}\rangle &\sim &\sqrt{\frac{a_0}{b_0}}\sum_k \frac{e^{ikx}}{(2(1-\cos k))^{B/4}}\notag\\
&\approx & \frac{2\pi}{L} \sqrt{\frac{a_0}{b_0}}\frac{2\cos k_0 x}{(2(1-\cos k_0))^{B/4}}\notag\\
&\approx & \frac{2\pi}{L} \sqrt{\frac{a_0}{b_0}}\frac{2\cos k_0 x}{k_0^{B/2}}\notag\\
&=&2\sqrt{\frac{a_0}{b_0}} \left(\frac{L}{\pi}\right)^{B/2-1}\times\left(2-\frac{\pi^2x^2}{L^2}\right).
\label{Ux0}
\end{eqnarray}
This result holds accurately for $B> 4$. To understand why, note that we have kept only the most divergent $k=k_0$ term in line 2, which corresponds to a momentum width $\Delta k = 2\pi/L$. The largest omitted terms scale like $(L/(3\pi))^{B/2}=(L/\pi)^{B/2}3^{-B/2}$, and can safely be discarded when $3^{-B/2}\ll 1$, i.e. when $B\approx 4$ or larger. Indeed, in Fig.~\ref{figa:fouriercompare}a, it was seen that $U_x$ follows an approximately parabolic profile for $B>4$. Cases with larger $B$ have similar profiles, except that they scale with $L^{B/2-1}$. 

\subsubsection{Rigorous characterization of divergence of 2-site propagators}

We next turn to a more rigorous derivation of the scaling of $U_x$ with $L$. It holds for generic $B$, and in particular allows us to characterize the special case of $B=2$, as well as $B<2$ cases. Not making any approximations, except that $L$ sets the lower IR limit~\footnote{This IR cutoff of the continuous momentum integral introduces some discrepancy with the discrete lattice momentum sum, but the overall scaling behavior for $x\ll L$ is preserved.}, we have (defining $k'=k/2$)
\begin{eqnarray}
U_x=-2\langle c^\dagger_{x,+}c_{0,-}\rangle &= &\sqrt{\frac{a_0}{b_0}}\int_{\pi/L}^\pi \frac{e^{ikx}dk}{(2(1-\cos k))^{B/4}}+\text{c.c}\notag\\
&=&\frac1{2^{B/2-1}}\sqrt{\frac{a_0}{b_0}}\int_{\pi/L}^\pi \frac{\cos kx \,dk}{((1-\cos k)/2)^{B/4}}\notag\\
&=&\frac1{2^{B/2-2}}\sqrt{\frac{a_0}{b_0}}\int_{\pi/2L}^{\pi/2} \frac{\cos 2k'x \,dk'}{\sin^{B/2}k'}\notag\\
&=&\frac1{2^{B/2-2}}\sqrt{\frac{a_0}{b_0}}\int_{\pi/2L}^{\pi/2}\left[\frac{1}{\sin^{B/2}k'}- \frac{2\sin^2k'x }{\sin^{B/2}k'}\right]dk'.
\label{Uxfull}
\end{eqnarray}
In particular, $x$ only enters the second term in the square parentheses, which is positive and vanishes when $x=0$ i.e. for the on-site inter-sublattice correlator. Hence $U_x$ is maximal for $x=0$, with its divergence with $L$ obtainable from the first term alone. In general, its scaling can be analytically derived in terms of a Hypergeometric function:
\begin{eqnarray}
U_0=-2\langle c^\dagger_{0,+}c_{0,-}\rangle &\sim &
-\frac1{2^{B/2-2}}\sqrt{\frac{a_0}{b_0}}\int\frac{dk'}{\sin^{B/2}k'}|_{k'=\pi/2L}\notag\\
&=&4\, _2F_1\left[\frac1{2},\frac1{2}-\frac{B}{4},\frac{3}{2}-\frac{B}{4},\sin^2 \frac{\pi}{2L}\right]\sqrt{\frac{a_0}{b_0}}\left(2\sin\frac{\pi}{2L}\right)^{1-B/2}\notag\\
&= & \frac{4}{B-2}\sqrt{\frac{a_0}{b_0}} \left(\frac{L}{\pi}\right)^{B/2-1}\left(1+\frac{B(B-2)}{3(B-6)}\frac{L^2}{\pi^2}+\frac{B(B-2)(5B+4)}{90(B-10)}\frac{L^4}{\pi^4}+...\right)
\label{Uxfull2}
\end{eqnarray}
whose leading order power-law divergence behavior agrees up to a constant with Eq.~\ref{Ux0}, which was obtained through hand-wavy arguments. Note that the above expansion converges only for $B\,(\text{mod}\,4)= 0$. A more careful analysis reveals logarithmic divergence behavior or non-universal constants the other cases, and subleading contributions for all cases:
\begin{itemize}
\item $B=1$: \begin{equation} U_0|_{B=1}\sim  \sqrt{\frac{32a_0}{b_0}}F\left(\frac{\pi(L-1)}{4L},2\right)= \sqrt{\frac{a_0}{b_0}}\left(7.4163-4\sqrt{\frac{\pi}{L}}+...\right)
\label{B1}
\end{equation}
\item $B=2$: \begin{equation} U_0|_{B=2}\sim   2\sqrt{\frac{a_0}{b_0}}\log\cot\frac{\pi}{4L}= 2\sqrt{\frac{a_0}{b_0}}\left(\log\frac{4}{\pi}+\log L+...\right)
\end{equation}
\item $B=3$: \begin{equation} U_0|_{B=3}\sim  \sqrt{\frac{8a_0}{b_0}}\left(\frac{\cos\frac{\pi}{2L}}{\sqrt{\sin\frac{\pi}{2L}}}-E\left(\frac{\pi(L-1)}{4L},2\right)\right)= \sqrt{\frac{a_0}{b_0}}\left(4\sqrt{\frac{\pi}{L}}-1.6944+...\right)
\end{equation}
\item $B=4$: \begin{equation} U_0|_{B=4}\sim  \sqrt{\frac{a_0}{b_0}}\cot\frac{\pi}{2L}= \sqrt{\frac{a_0}{b_0}}\left(\frac{2L}{\pi}-\frac{\pi}{6L}+...\right)
\end{equation}
\item $B=6$: \begin{equation} U_0|_{B=6}\sim  \frac1{4}\sqrt{\frac{a_0}{b_0}}\left(\log\cot\frac{\pi}{4L}+\frac{\cos\frac{\pi}{2L}}{\sin^2\frac{\pi}{2L}}\right)= \sqrt{\frac{a_0}{b_0}}\left(\frac{L^2}{\pi^2}-\frac1{24}+...\right)
\label{B6}
\end{equation}
\end{itemize}
Here $F(\phi,m)=\int_0^\phi d\theta/\sqrt{1-m\sin^2\theta}$ and $F(\phi,m)=\int_0^\phi d\theta \sqrt{1-m\sin^2\theta}$ are the Elliptic integrals. As evident from Fig \ref{figa:fouriercompare}a, odd $B$ cases can be realized with the ansatz $b(k)=b_0(1-e^{ik})^B$, with the caveat that $U_x=0$ for all even $x$. 

In particular, the marginal $B=2$ case, which can be realized commonly in lattices with nearest neighbor couplings, contains two-pt functions $-U_0/2$ that diverges logarithmically with $L$. The $B=1$ case, which can be realized with $b(k)=b_0(1-e^{ik})$  that also contains up to only nearest neighbor couplings, is also noteworthy in that $U_0$ saturates to a constant $4\sqrt{\frac{a_0}{b_0}}F(\pi/2,1/2)=7.4163\sqrt{\frac{a_0}{b_0}}$ as $L\rightarrow \infty$. While this lack of divergence is special among all these cases containing an EP, it is typical among non-EP gapless scenarios, where the on-site inter-sublattice propagator should not grow with $L$.

In general, the 2-site propagator $U_x$ scales with $x$ in the same way as it scales with $L$, since these are the only length scales present, except when $x$ is comparable with $L$. This can be explicitly demonstrated for some simple but important cases in the $L\rightarrow \infty$ limit. For $B=2$, we have from Eq.~\ref{Uxfull}
\begin{eqnarray}
U_x|_{B=2}&=& U_0|_{B=2}-\frac1{2^{1-2}}\sqrt{\frac{a_0}{b_0}}\int_{0}^{\pi/2} \frac{2\sin^2k'x }{\sin k'}dk'\notag\\
&= & U_0|_{B=2}-4\sqrt{\frac{a_0}{b_0}}\sum_{j=1}^x\frac{\cos jk}{2j-1}|_{k'=0}^{k'=\pi/2}\notag\\
&= & U_0|_{B=2}-4\sqrt{\frac{a_0}{b_0}}\sum_{j=1}^x\frac{1}{2j-1}\notag\\
&\sim & U_0|_{B=2}-2\sqrt{\frac{a_0}{b_0}}\left(\log x+\gamma+\log 4\right)\notag\\
&= & 2\sqrt{\frac{a_0}{b_0}}\left(\log \left(\frac1{4x}\cot\frac{\pi}{4L}\right)-\gamma\right)\notag\\
&= & 2\sqrt{\frac{a_0}{b_0}}\left(\log \frac{L}{\pi x}-\gamma+...\right)
\end{eqnarray}
where $\gamma$ is the Euler-Mascheroni constant from the approximation in line 4, and the last line contains subleading terms vanishing as $1/x$, $1/L$ or faster. 

For the $B=4$ case, we also have
\begin{eqnarray}
U_x|_{B=4}&=& U_0|_{B=4}-\frac1{2^{2-2}}\sqrt{\frac{a_0}{b_0}}\int_{0}^{\pi/2} \frac{2\sin^2k'x }{\sin^2 k'}dk'\notag\\
&= & U_0|_{B=4}-2\sqrt{\frac{a_0}{b_0}}\left(k'x+[\sin...]\right)|_{k'=0}^{k'=\pi/2}\notag\\
&=& U_0|_{B=4}-\pi x \notag\\
&=&  2\sqrt{\frac{a_0}{b_0}}\left(\frac{L}{\pi}-\pi x\right)
\end{eqnarray}
i.e. that $U_x|_{B=4}$ depends linearly on both $L$ and $x$, in qualitative agreement with Fig.~\ref{figa:fouriercompare}a~\footnote{For a finite truncated region, the exact linear relationship between $L$ and $x$ depends on the UV regularization i.e. $U_x|_{B=4}\propto L-2x$ in Fig.~2 rather than $L-\pi^2x$. }.

\subsection{Emergence of EB eigenstates}

EB eigenstates emerge when the eigenstate is forced to have compact support, and can arise from even a single boundary. Setting the boundary to be at $x=0$, we return to Eq.~\ref{Im2P2} which shows that Fourier coefficients $D_j$ are relatively small and rapidly decaying, but Fourier coefficients $U_j$ are diverging like $\sim L^{B/2}$ and slowly decaying with $j$. These turn out to be the exact conditions needed to support Exceptional Boundary (EB) eigenstates of $\bar P$. These special eigenstates have eigenvalues $p_\pm=\frac1{2}(1\pm\sqrt{1+\lambda})$ that are well-separated from the other eigenvalues which are clustered around $0$ and $1$. Focusing on the upper $+$ sublattice (the $-$ sublattice eigenequation contains equivalent information), the EB eigenvalue $\lambda$ appears in the eigenequation
%Suppose $|\phi\rangle$ is an eigenstate of $\bar P$. Then it is also an eigenstate of $\bar P^2$ and hence $\Lambda=4(\bar P^2-\bar P)=(\mathbb{I}-2\bar P)^2-\mathbb{I}$. Since we already know from Eq.~\ref{Im2P2} that the latter decouples into two sublattice sectors, we decompose $|\phi\rangle=|\phi_1\rangle\oplus|\phi_2\rangle$, such that $\Lambda_{jj}|\phi_j\rangle=\lambda|\phi_j\rangle$, with $\Lambda_{jj}$ being the block of $\Lambda$ representing sublattice $j$. Focusing on $j=1$ for convenience, we have \red{check!}
\begin{equation}
\sum_{x_1}\langle x_1|\Lambda_+|x_2\rangle\phi_+(x_1)=
\sum_{x_1}\sum_{\substack{x_u\leq x_1,\\ x_d\leq x_2}} [U_{-x_u}D_{x_d}\delta_{x_u+x_d,x_1-x_2}-\delta_{x_1,x_2}]\phi_+(x_1)=\lambda \phi_+(x_2).
\label{eigeneq}
\end{equation}
More explicitly, making use of the fact that $D(k)U(k)=1$ i.e. $\sum_{x} D_{x+\Delta} U_{-x}=\delta_{\Delta,0}$, the $x_1$-th column of the $L\times L$ matrix $\Lambda_+=\bar U\bar D-\mathbb{I}$ takes the form
\begin{equation}
-|c_{x_1}\rangle=D_{x_1}\left(\begin{matrix}
U_{-1}\\ U_{-2} \\ U_{-3}\\\vdots
\end{matrix}\right)+
D_{x_1+1}\left(\begin{matrix}
U_{-2}\\ U_{-3} \\ U_{-4}\\\vdots
\end{matrix}\right)+... = \sum_{j=0}D_{x_1+j}\left(\begin{matrix}
U_{-1-j}\\ U_{-2-j} \\ U_{-3-j}\\\vdots
\end{matrix}\right).
\label{ca}
\end{equation}
Analogously, the $x_2$-th row of $\langle x_1|\Lambda_+|x_2\rangle=\langle x_1|(\bar U\bar D-\mathbb{I})|x_2\rangle$ takes the form
\begin{equation}
-\langle r_{x_2}|=U_{-x_2}\left(\begin{matrix}
D_{1}\\ D_{2} \\ D_{3}\\\vdots
\end{matrix}\right)^T+
U_{-x_2+1}\left(\begin{matrix}
D_{2}\\ D_{3} \\ D_{4}\\\vdots
\end{matrix}\right)^T+... = \sum_{j=0}U_{-x_2-j}\left(\begin{matrix}
D_{1+j}\\ D_{2+j} \\ D_{3+j}\\\vdots
\end{matrix}\right)^T.
\label{ra}
\end{equation}
Importantly, the slow decay of the $U_{-j}$ coefficients, which are already divergently scaling as $\sim L^{B/2}$, ensures that the matrix $\Lambda_+=\bar U\bar D-\mathbb{I}$ is dominated by its first  slowly-decaying column. The subsequent columns contain much smaller elements due to the rapid decay of the $D_j$ coefficients. Meanwhile, the situation with the rows is \emph{not} analogous. Due to the rapid decay of the $D_j$ coefficients, none of the rows will contain large, slowly-decaying elements, perhaps with the exception of the first. With a dominant first column $|c_1\rangle$ and subdominant first row $\langle r_1|$, the eigenequation Eq.~\ref{eigeneq} can be approximated as
\begin{comment}
\begin{equation}
\Lambda_+\left(\begin{matrix}
\phi_+(1)\\ \phi_+(2) \\ \phi_+(3)\\\vdots
\end{matrix}\right)\approx \sum_{j,j'=0}D_{1+j}U_{-1-j'}\left(\begin{matrix}
U_{-1-j}\\ U_{-2-j} \\ U_{-3-j}\\\vdots
\end{matrix}\right)(D_{1+j'},D_{2+j'},D_{3+j'},...)\left(\begin{matrix}
\phi_+(1)\\ \phi_+(2) \\ \phi_+(3)\\\vdots
\end{matrix}\right)=\lambda\left(\begin{matrix}
\phi_+(1)\\ \phi_+(2) \\ \phi_+(3)\\\vdots
\end{matrix}\right)
\end{equation}
or, in more compact notation,
\end{comment}
\begin{equation}
\Lambda_+|\phi_+\rangle \approx \left[|c_1\rangle \langle 1|+|1\rangle\langle r_1|-\langle 1|c_1\rangle|1\rangle\langle 1|\right]|\phi_+\rangle = \lambda|\phi_+\rangle. 
\end{equation}
The first two terms gives the dominant column and subdominant row respectively, while the third term subtracts the $\langle 1|\Lambda_+|1\rangle =\langle 1|c_1\rangle$ matrix contribution that appears in both $\langle r_1|$ and $|c_1\rangle$. Now, because the first term $|c_1\rangle\langle 1|$ is the dominant contribution, the eigenvector $|\phi_\rangle$ must be approximately parallel to $|c_1\rangle$. Substituting $|\phi_+\rangle \approx |c_1\rangle$ and left multiplying the above by $\langle 1|$, we obtain
\begin{equation}
\langle 1|c_1\rangle ^2 + \langle 1|1\rangle \langle r_1|c_1\rangle -\langle 1|c_1\rangle\langle 1|1\rangle\langle 1|c_1\rangle \approx \lambda \langle 1|c_1\rangle
\end{equation}
i.e.
\begin{eqnarray}
\lambda \approx \frac{\langle r_1|c_1\rangle}{\langle 1|c_1\rangle}&=& \frac{\langle 1|\Lambda^2_+|1\rangle}{\langle 1|\Lambda_+|1\rangle}\label{firsta}\\
&=& \frac{\langle 1|(\bar U\bar D\bar U\bar D-2\bar U\bar D+\mathbb{I})|1\rangle}{\langle 1|(\bar U\bar D-\mathbb{I})|1\rangle}\notag\\
&=&-\frac{\sum_{j,j'=0}D_{1+j}U_{-1-j'}(D_{1+j'},D_{2+j'},D_{3+j'},...)}{\sum_{j=0}D_{1+j}U_{-1-j}}\cdot\left(\begin{matrix}
U_{-1-j}\\ U_{-2-j} \\ U_{-3-j}\\\vdots
\end{matrix}\right)\notag\\
&=&-\frac{\sum_{j,j',j''=0}D_{1+j}U_{-1-j'}D_{1+j''+j'}U_{-1-j-j''}}{\sum_{j=0}D_{1+j}U_{-1-j}}.
\label{lambdaa}
\end{eqnarray}
The first line (Eq.~\ref{firsta}) follows from the definition of $\langle 1|$, $|c_1\rangle$ as the first row and column of $\Lambda_+$, which can be expressed in terms of the final explicit expressions upon making use of the simplifications used in Eqs.~\ref{ca} and \ref{ra}. 

Eq.~\ref{firsta} expresses the special EB eigenvalue $\lambda$ of $\Lambda_+$ in terms of the corner matrix elements of $\Lambda_+$ and $\Lambda_+^2$, that can be easily computed \emph{without} further explicit diagonalization. Although we have made various approximations along the way, Eq.~\ref{firsta} or \ref{lambdaa} is highly accurate for a huge range in parameter space, even when the EP is not too singular ($B\geq 2$). Note that this $\lambda$ eigenvalue is unique and the other eigenvalues, which constitute the majority, are still very close to $0$. 

\begin{figure}
\subfloat[]{\includegraphics[width=.41\linewidth]{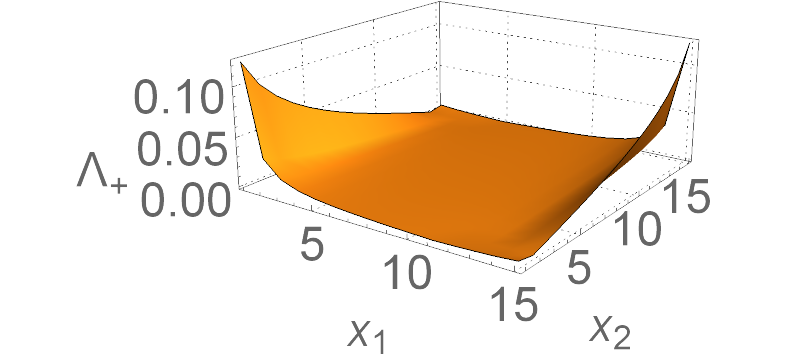}}
\subfloat[]{\includegraphics[width=.35\linewidth]{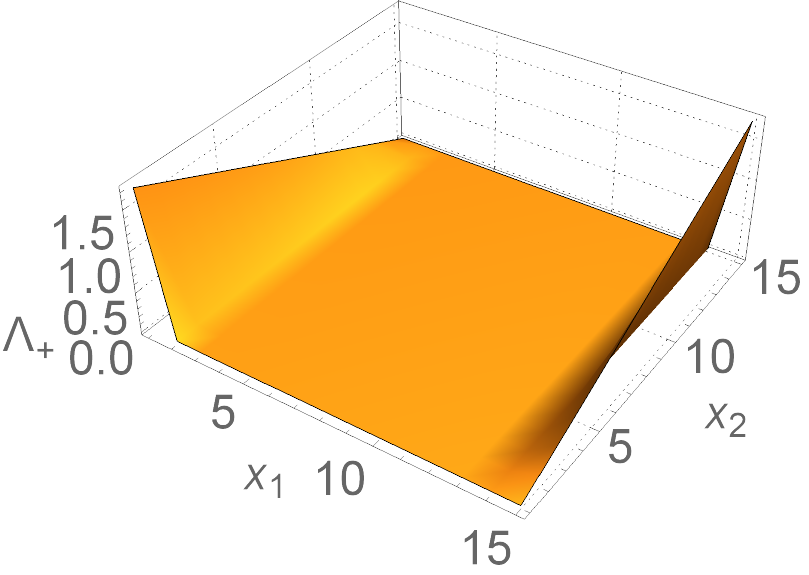}}\\
\subfloat[]{\includegraphics[width=.39\linewidth]{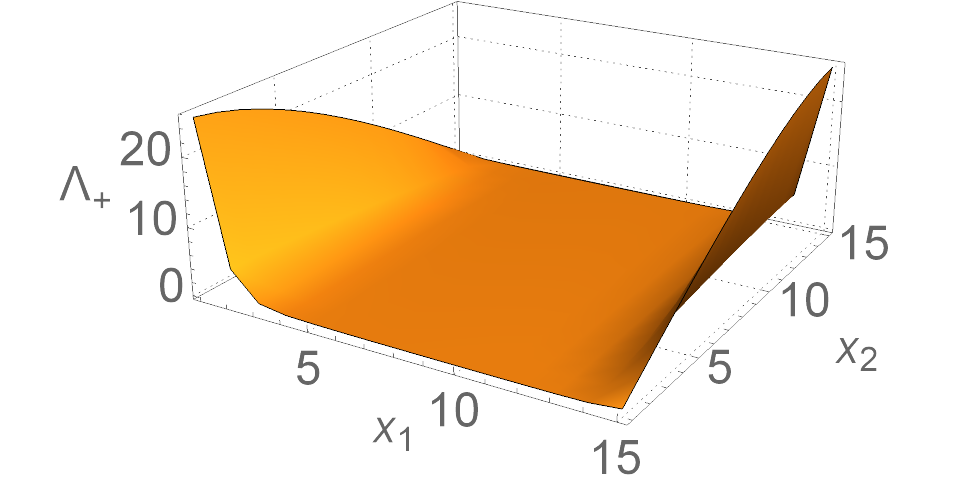}}
\subfloat[]{\includegraphics[width=.39\linewidth]{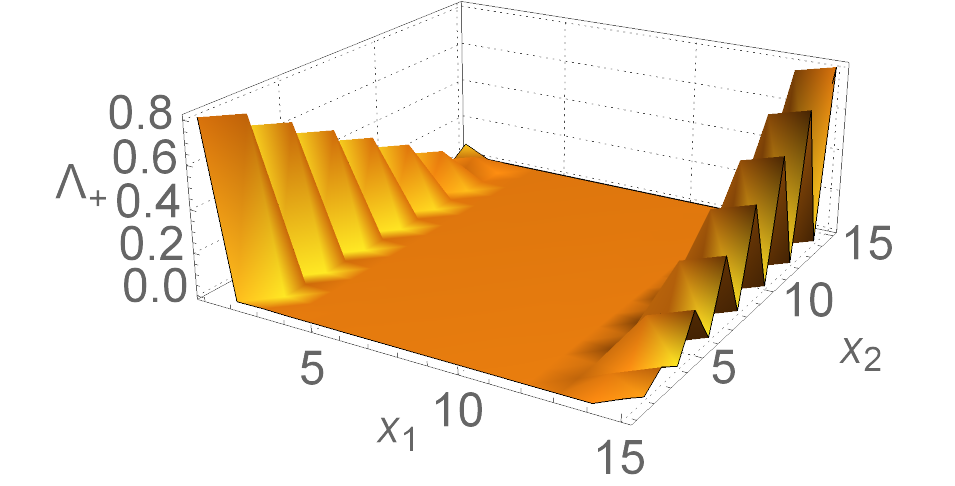}}
\caption{
Profiles of the matrix elements of $\Lambda_+$ for (a) $B=2$, (b) $B=4$, (c) $B=6$ order EPs with $b(k)=(2(1-\cos k))^{B/2}$ and $a_0=10^3$. While most elements are close to $0$, similar to the Hermitian case where $\Lambda_+$ vanishes, the first column matrix elements are anomalously large in our cases. Their characteristic concave, linear and quadratic spatial profiles dominate the effects of $\Lambda_+$, and in fact determine the profiles of their corresponding EB eigenstates. (d) showcases the profile of an alternative $B=4$ EP implementation based on $b(k)=(2i\sin k_x)^4$, which is similar to (b) except for odd/even effects.
}
\label{figa:matrix}
\end{figure}

%\subsection{Profile of EB eigenstates}

The fact that the EB eigenstate $|\phi_+\rangle$ and $|c_1\rangle$, the first column of $\Lambda_+$, are almost parallel implies that the EB eigenstate profile is largely determined by the Fourier components of $D(k)$ and $U(k)$ via Eq.~\ref{ca}. For more singular cases in particular, $\phi_+\rangle$ and $|c_1\rangle$ are almost exactly parallel, and the components of $U_j$ accurately determine the EB state. As plotted in Fig.~\ref{figa:matrix},  the profile of the first column $|c_1\rangle$ resemble the profiles of the EB states in Fig.~2a of the main text. As $B$ increases, the size of the elements in $|c_1\rangle$ increases strongly, leading to more well-defined EB modes.
%\begin{figure}
%\includegraphics[width=1\linewidth]{????}
%\caption{\red{
%} }
%\label{figa:profiles}
%\end{figure}

%\red{what else needs to be plotted?}

\subsection{Scaling of EB eigenvalues and entanglement entropy}

From previous subsections, we already know that the Fourier coefficients $U_j$ diverge like $L^{B/2-1}$ for $B>2$, but that the first few Fourier coefficients $D_j$ remain nonzero but fixed. Hence from Eq.~\ref{lambdaa}, we expect $\lambda$ to scale for $B>2$ like
\begin{equation}
\lambda\sim \lambda_0 L^{B/2-1} = -\frac{\sum_{j,j',j''=0}D_{1+j}u_{-1-j'}D_{1+j''+j'}u_{-1-j-j''}}{\sum_{j=0}D_{1+j}u_{-1-j}}L^{B/2-1}
\label{scal}
\end{equation}
where $\lambda_0$ is a constant determined by the quotient above, and $u_j = U_j/L^{B/2-1}$ are the Fourier coefficients with $L^{B/2-1}$ divergence factored out. Likewise, for $B=2$, we have $\lambda|_{B=2} \sim \lambda_0\log L$. For $B<2$, the first column of $\Lambda_+$ are not dominant and the derivations leading to Eq.~\ref{lambdaa} are not applicable. As such, a well-defined $\lambda$ and hence EB states do not exist [Fig.~2a of the main text].

For each boundary, the EE contribution from the EB state is, for $B>2$,
\begin{eqnarray}
S_\text{EB} &=& -\text{Tr}_\text{EB}[\bar P\log \bar P +(\mathbb{I}-\bar P)\log(\mathbb{I}-\bar P) ]\notag\\
&=& -\sum_\pm\left[ \frac{1\pm\sqrt{1+\lambda}}{2}\log\frac{1\pm\sqrt{1+\lambda}}{2}+\frac{1\mp\sqrt{1+\lambda}}{2}\log\frac{1\mp\sqrt{1+\lambda}}{2}\right]\notag\\
&\sim &\log 4-2-2\log\sqrt{1+\lambda}\notag\\
&\sim & \log 4-2 -\log \lambda_0 - \frac{B-2}{2}\log L.
\label{SEB}
\end{eqnarray} 
On the first line, the trace is only taken over the possible EB states. The first $\sim$ refers to the asymptotic expansion of the 2nd line at large $\lambda$, while the second $\sim$ pertains to the large $L$ behavior of $\lambda$. We see that a divergence of positive power $B$ always tends to suppress the EE, in fact contributing a quantized negative coefficient to $\log L$. In the presence of other competing terms i.e. $\sigma_z$ in the Hamiltonian, the EB modes become less protected by the defectiveness, and varying amounts of positive $\sim \log L$ contributions will also be present, as presented in a later section. The marginal $B=2$ case has $\lambda$ increasing with $L$ logarithmically slowly, resulting in $S_{EB}\sim -\log\log L +\text{const.}$ at very large $L$. This weak double logarithmic increase can be easily overshadowed by other $\sim \log L$ or larger contributions should any competing term be present. Indeed, it was numerically shown that $S\sim -2/3\log L$ for $B=2$ in Fig.~3a of the main text, as also justified by CFT for a related model in Ref.~\cite{chang2020entanglement}.

In generic entanglement geometries, the above expression will be multiplied by the number of entanglement cuts i.e. 1 (or 2) cuts for a single entanglement subsystem in a 1D  OBC (or PBC) system.

%\red{comment on adding $\gamma(k)$}
\begin{comment}
\begin{figure}
\subfloat[]{\includegraphics[width=.44\linewidth]{fig_gamma_B=2.pdf}}
$\qquad$
\subfloat[]{\includegraphics[width=.44\linewidth]{fig_gamma_B=4.pdf}}
\caption{\red{
may bring to main text... }
}
\label{fig:EEgamma}
\end{figure}
\red{note that $a_0=10^3$ throughout, unless otherwise stated}
\end{comment}

\subsection{Secondary EB states and beyond}
\begin{figure}
\subfloat[]{\includegraphics[width=.33\linewidth]{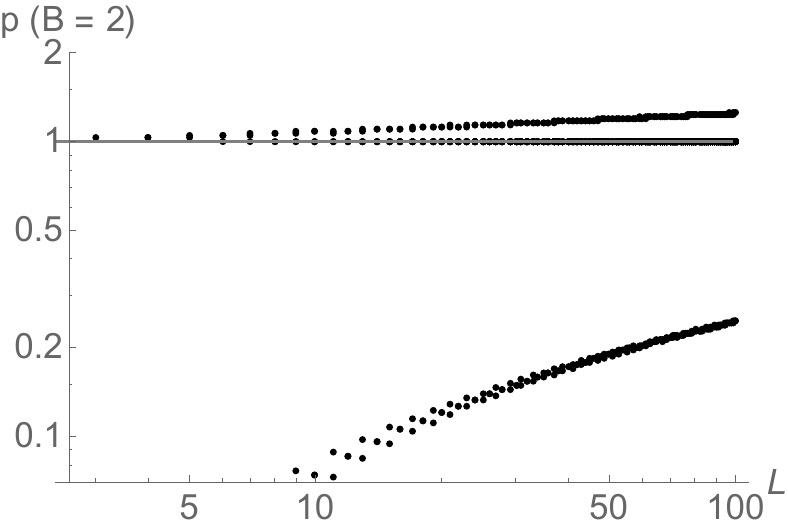}}
\subfloat[]{\includegraphics[width=.33\linewidth]{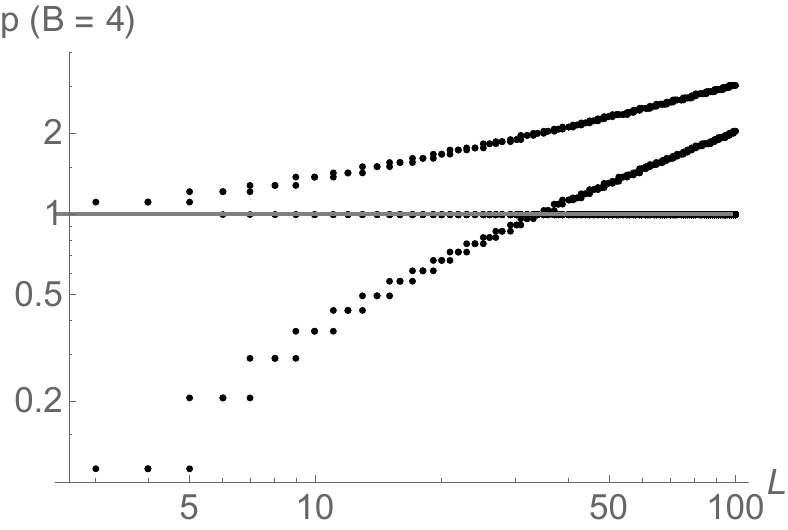}}
\subfloat[]{\includegraphics[width=.32\linewidth]{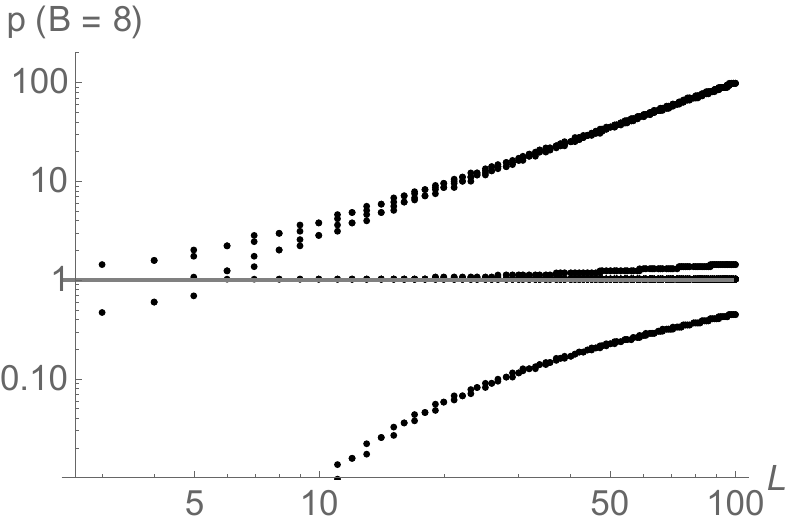}}\\
\subfloat[]{\includegraphics[width=.33\linewidth]{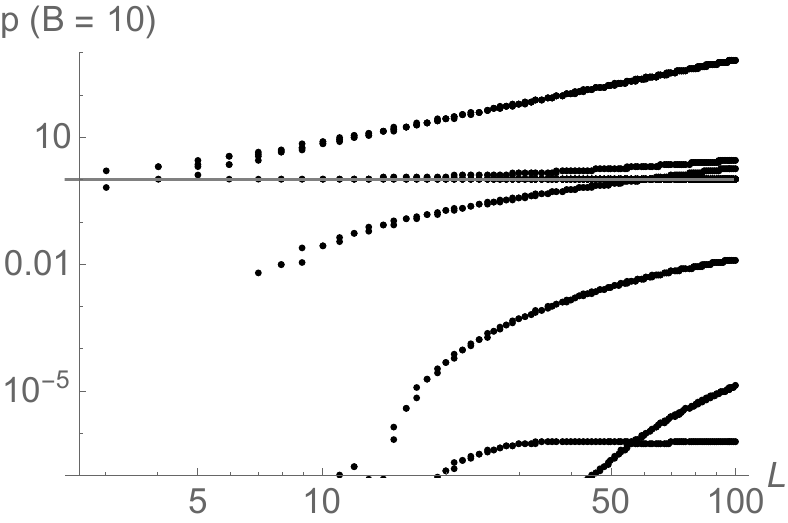}}
\subfloat[]{\includegraphics[width=.33\linewidth]{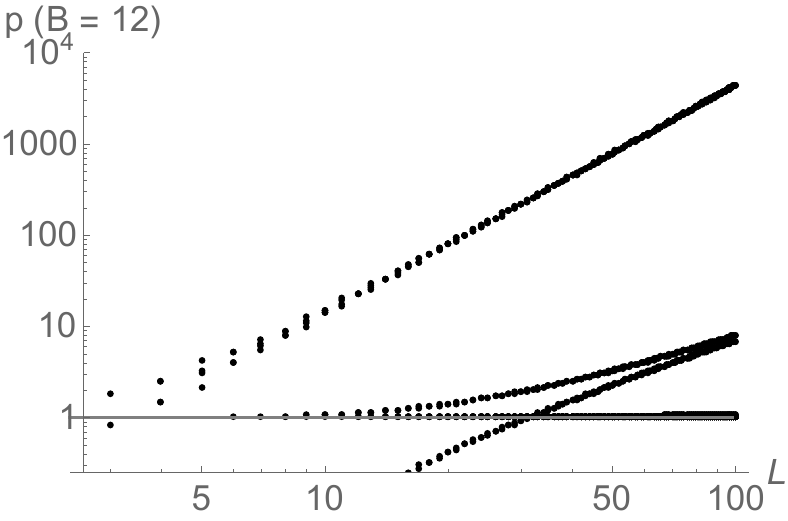}}
\subfloat[]{\includegraphics[width=.32\linewidth]{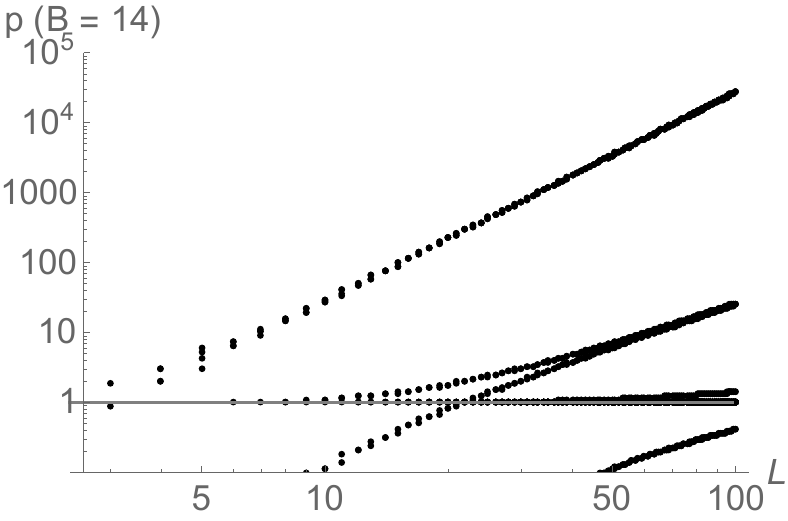}}\\
\subfloat[]{\includegraphics[width=.31\linewidth]{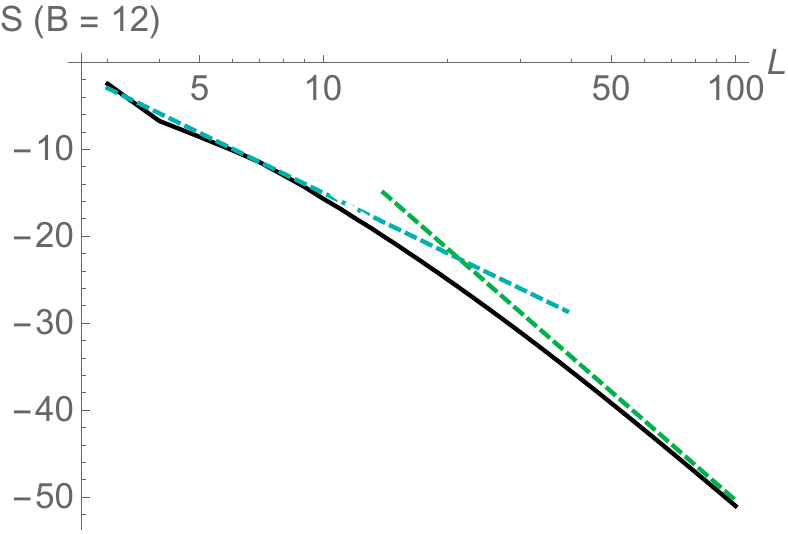}}
\subfloat[]{\includegraphics[width=.32\linewidth]{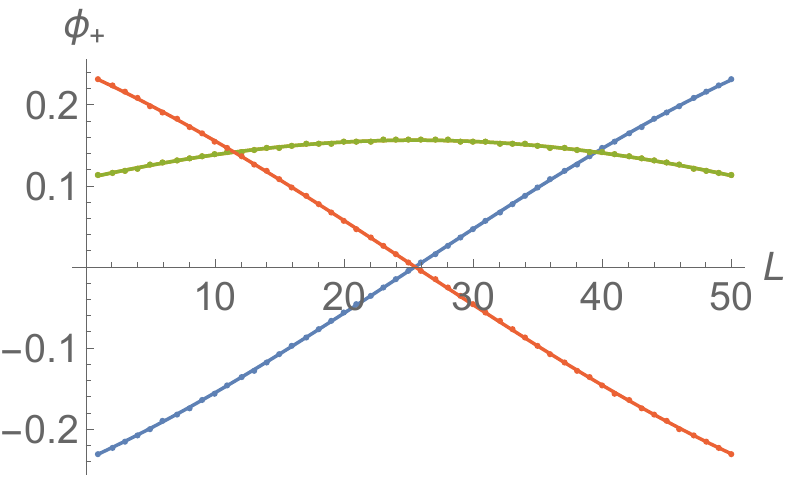}}
\subfloat[]{\includegraphics[width=.32\linewidth]{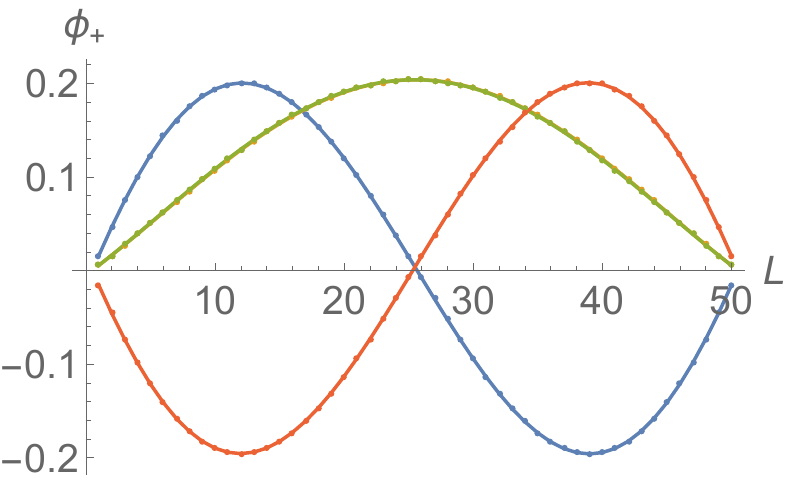}}
\caption{(a-f) Magnitudes of the eigenvalues $p$ of $\bar P$ as a function of $L$, plotted on log-log scales for $B=2,4,8,10,12,14$ respectively for the ansatz model Eq.~\ref{abk} with $a_0=10^3$. While only one pair of $|p|>1$ eigensolutions (EB states) exist at $B=2$ and $B=4$, a secondary set of EB state eigensolutions emerge for $B\geq 8$. At $B=14$, a tertiary set of EB states also appear. For the $B=12$ case, the emergence of secondary EB states at $L\approx 30$ (e) corresponds to the crossover of the EE $S$ behavior. Below $L\approx30$, $S\sim -(B-2)\log L=-10\log L$ [Eq.~\ref{SEB}] (cyan dashed), but above $L\approx30$, $S\sim -(B-2)\log L - (B-4)\log L=-18\log L$ (green dashed). Primary (h) and secondary (i) EB states differ markedly in their spatial profiles.
}
\label{figa:secondary}
\end{figure}

In the case of very strong divergences i.e. large $B$, $\Lambda$ contains not just one set of strongly nonvanishing eigenvalues, but also secondary, tertiary etc. sets as well. They arise because a very large $B$ leads to very large elements in the first few columns of $\Lambda_+$, not just the first column as for the EB states described previously. This can be understood from Eq.~\ref{ca}. Large $B$ leads to divergent values of $U_x$ [Eq.~\ref{Ux0}], and more importantly more nonlocal $D_{x_1+j}$ in Eq.~\ref{ca} which leads to more slowly decaying columns in $\Lambda_+$. Generalizing the mechanism involved in the derivation of Eq.~\ref{lambdaa}, we empirically find additional \emph{secondary} EB states that appear at sufficiently large $L$, scaling as $\sim L^{B/2-2}$ instead of $\sim L^{B/2-1}$. 

The emergence of these additional EB states is studied in Fig.~\ref{figa:secondary}. At small $B$, i.e. $B=2$, there is only one set of weak EB state with eigenvalue $p>1$. The other states with $p\in[0,1]$ behave like ordinary truncated projector eigenstates. At $B=4$, the corresponding branch of eigenvalues $1-p$ satisfies $|1-p|>1$ for larger values of $L$, but still have the same origin as the $p$ eigenvalues. But at larger $B$, i.e. at $B=8$, a new distinct branch of $p>1$ eigenvalues emerge. These secondary $p$ eigenvalues can become very large too for even larger $B$, albeit scaling like $\sim L^{B/2-2}$ instead of $\sim L^{B/2-1}$. This is verified in Fig.~\ref{figa:secondary}g, where different sets of EB states contribute additively to the EE. At $B=14$, we even observe the emergence of a set of \emph{tertiary} EB modes with $p>1$. Note that in general, the spatial profiles of secondary EB states result from complicated combinations of the profiles of the 2-site propagator [Fig.~\ref{figa:secondary}i], unlike those of ordinary (primary) EB states [Fig.~\ref{figa:secondary}h].

\section{Temporal evolution of the two-point function}

For a non-Hermitian system, the temporal two-point function with respect to a given operator $\varphi(t)$ is given by
\begin{equation}
\langle \varphi^\dagger(t)\tilde\varphi(0)\rangle = \langle \tilde\varphi(0)|P|\varphi(t)\rangle= \langle \tilde\varphi|Pe^{-iH_Ft/\hbar}|\varphi\rangle,
\end{equation}
where $H_F$ is the Hamiltonian for $t\geq 0$, $P$ is the projector onto the occupied Hilbert space and $|\varphi\rangle=|\varphi(0)\rangle = \varphi^\dagger(0)|0\rangle$.

\begin{figure}
\subfloat[]{\includegraphics[width=.26\linewidth]{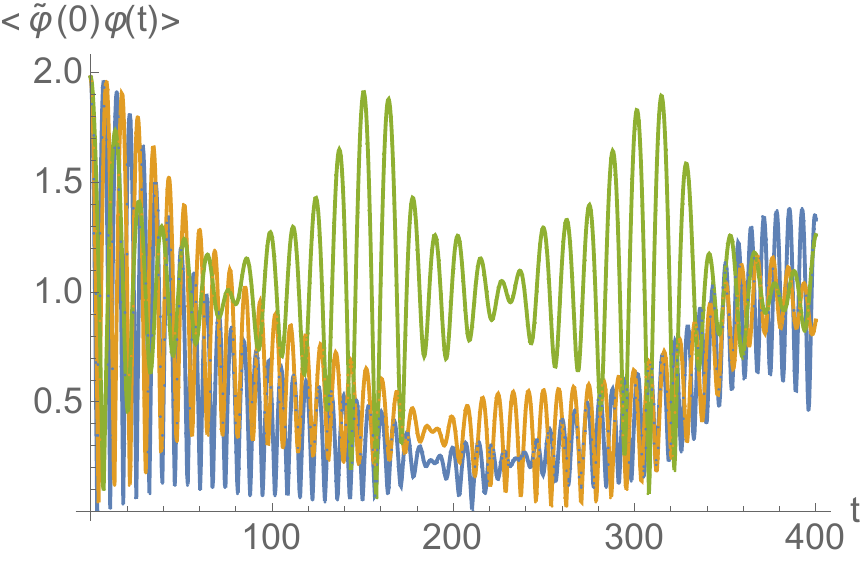}}
\subfloat[]{\includegraphics[width=.26\linewidth]{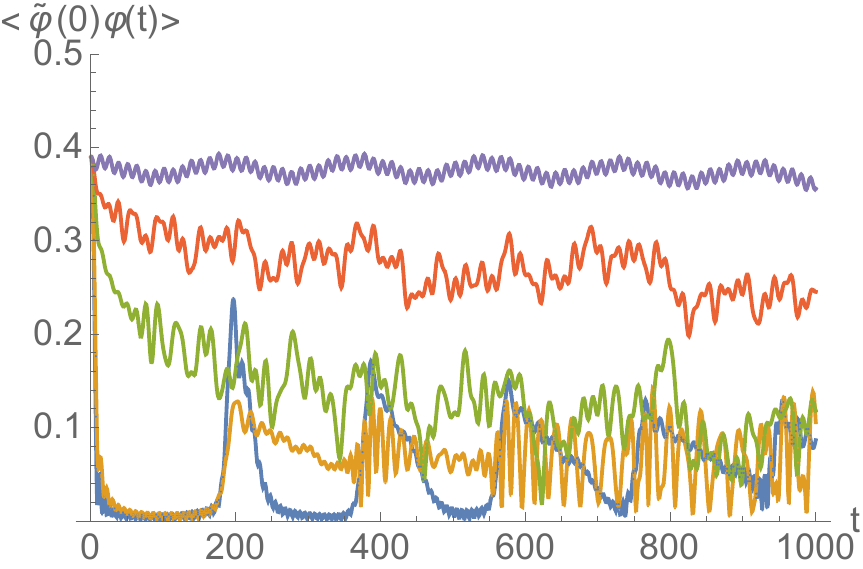}}
\subfloat[]{\includegraphics[width=.26\linewidth]{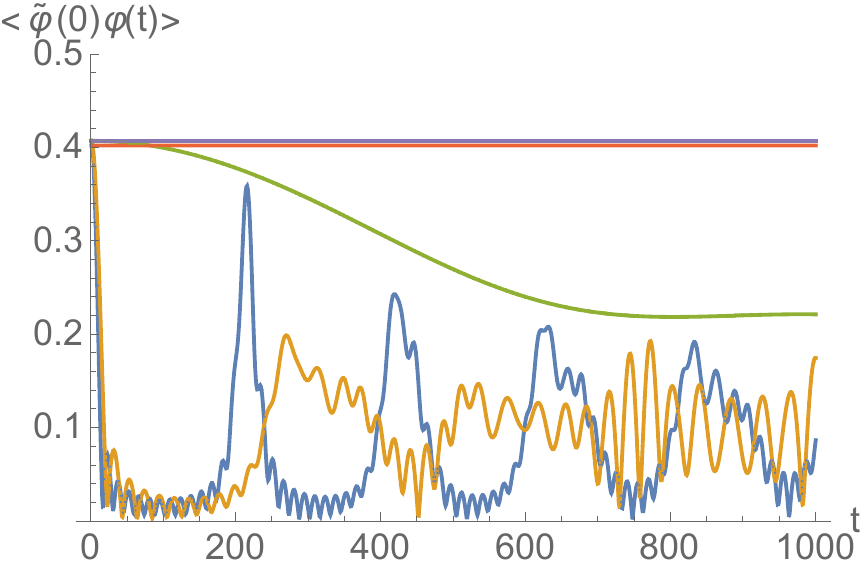}}
\subfloat[]{\includegraphics[width=.2\linewidth]{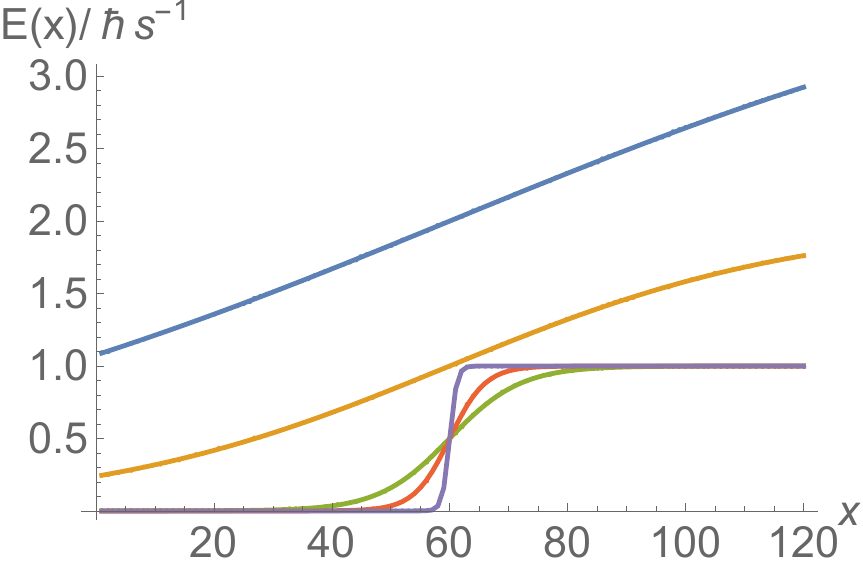}}
\caption{Time evolution of the 2-point function of a (a) $B=4$ EB state $\varphi(x)$ with $a_0=1$, (b) uniform state within the position interval $x\in [0,L/2]$ and (c) uniform intial state within $x\in [0,L/4]$. The various colored curves correspond to evolutions due to different Hamiltonians $H_F$ [Eq.~\ref{HFapp}] with real space profiles as color coded in (d). Specifically, the potential well slopes are given by $\alpha=1,2,10,20,100$, $E_0=1,1,1,2,3$ (purple,red,green,orange,blue), and $x_0=L/2=60$ for all cases. In (a), the 2-point function slightly decays and oscillates for the EB state, but often remains elevated above unity. But the 2-point function take on much smaller values for other generic states, as in (b) and (c), typically with an initial decay followed by oscillations that eventually degenerate into random scrambling.}
\label{figa:corrEx}
\end{figure}

To be consistent with probabilistic interpretations, the bra state $\langle\tilde\varphi|$ is taken to be the biorthogonal conjugate of the ket $|\varphi\rangle$. To see why, we briefly review the concepts biorthogonal bases. Suppose that $\langle \tilde\phi_i|$ and $|\phi_i\rangle$  are corresponding left and right eigenstates of an operator $H$ i.e. 
\begin{subequations}
\begin{align}
H|\phi_i\rangle &= h_i|\phi_i\rangle,\\
\langle \tilde \phi_i|H&= h_i\langle\tilde\phi_i|.
\end{align}
\end{subequations}
Under a similarity basis transform $H\rightarrow VHV^{-1}$, which is not necessarily unitary, the above equations give us $|\phi_i\rangle\rightarrow V|\phi_i\rangle$ and $\langle \tilde\phi_i|\rightarrow \langle\tilde \phi_i|V^{-1}$. Since $VV^{-1}=V^{-1}V=\mathbb{I}$, biorthogonal expectations are independent of basis transformations. This would not have been true if the bra had been $\langle \phi_i|$, the Hermitian conjugate of $|\phi_i\rangle$, since $\langle \phi_i| \rightarrow \langle\phi_i|V^\dagger$ and the expectation of an arbitrary operator $\hat O$ will have transformed like $\langle\phi_i| \hat O|\phi_i\rangle \rightarrow \langle\phi_i| V^\dagger\hat O V|\phi_i\rangle $, which is not consistent with the similarity transformation $\hat O\rightarrow V\hat OV^{-1}$. In particular, normalization and orthonormality is preserved in the biorthogonal basis only: $\langle \tilde\phi_j|\phi_i\rangle \rightarrow \langle \tilde\phi_j|V^{-1}V|\phi_i\rangle=\langle \tilde\phi_j|\phi_i\rangle=\delta_{ij}$ but not $\langle \phi_j|\phi_i\rangle\rightarrow\langle \phi_j|V^\dagger V|\phi_i\rangle $.

Suppose that $|\varphi\rangle=\sum_i \chi_i|\phi_i\rangle$ and $\langle\tilde\varphi|=\sum_i \bar \chi_i\langle \tilde \phi_i|$, where $\chi_i=\langle \tilde \phi_i|\varphi\rangle$ and $\bar\chi_i=\langle \tilde \varphi|\phi_i\rangle$. The expectation of $H$ is
\begin{equation}
\langle H\rangle = \langle \tilde \varphi|H|\varphi\rangle=\sum_{ij}\chi_i\bar\chi_j \langle \tilde \phi_j|H|\phi_i\rangle = \sum_i h_i \bar\chi_i\chi_i.
\end{equation}
If $\langle\tilde \varphi|$ is defined such that $\bar \chi_i=\chi^*_i$, we arrive at the probabilistic expression
\begin{equation}
\langle H\rangle =  \sum_i h_i |\chi_i|^2
\end{equation}
where $\sum_i  |\chi_i|^2=1$ from biorthogonality. (In the Hermitian context, there is no distinction between usual orthogonality and biorthogonality because only unitary rotations $V$ are considered i.e. $V^\dagger V=\mathbb{I}$.)

At $t=0$, $\langle \varphi^\dagger(0)\tilde\varphi(0)\rangle=\langle \tilde\varphi|P|\varphi\rangle$ is the density expectation of the state $|\varphi\rangle=|\varphi(0)\rangle$. As shown in the main text, this density can be anomalously large when $|\varphi\rangle=|\phi\rangle$, an EB state. Interestingly, $\langle \varphi^\dagger(t)\tilde\varphi(0)\rangle$ continues to remain elevated at $t>0$ for $|\varphi\rangle$ an EB state, as long as the energy profile does not excessively erode the EB profile. As shown in Fig.~\ref{figa:corrEx}, the 2-point function generally decays initially before transiting into oscillatory behavior [Figs.~\ref{figa:corrEx}a-c]. For the EB  state [Fig.~\ref{figa:corrEx}a], whose density expectation is $\approx 2$, the 2-point function remains elevated and frequently exceeds unity. But for other states i.e. uniform profiles $\langle x|\varphi\rangle=\varphi(x)=\theta(L/2-x)$ [Figs.~\ref{figa:corrEx}b] or $\langle x|\varphi\rangle=\varphi(x)=\theta(L/4-x)$ [Figs.~\ref{figa:corrEx}c], the 2-point function starts at $\approx 0.4$, and decays further to smaller values. 

\begin{figure}
\subfloat[]{\includegraphics[width=.37\linewidth]{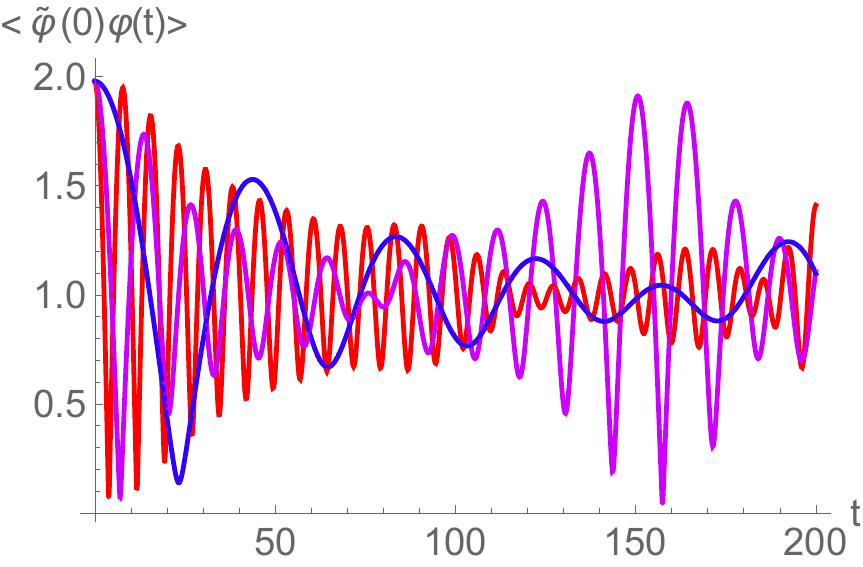}}
\subfloat[]{\includegraphics[width=.37\linewidth]{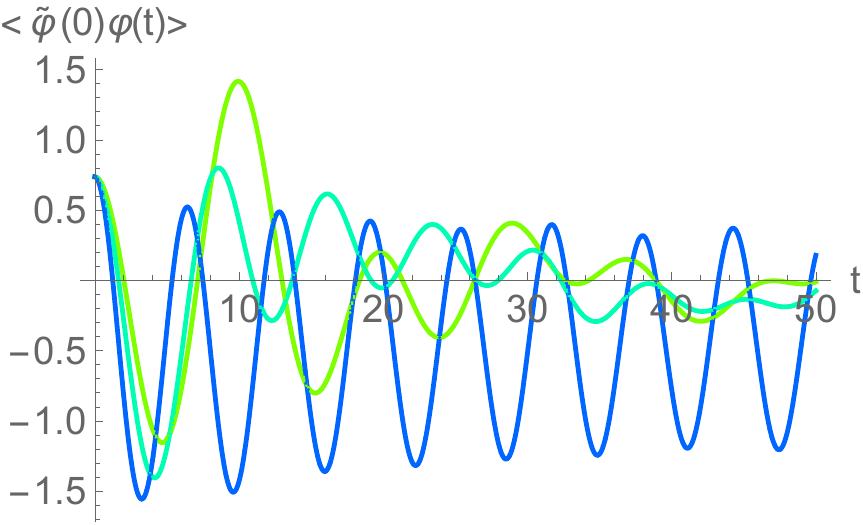}}
\caption{(a) Dependence of 2-point function evolution with potential wall position $x_0=25,30,35$ (red,magenta,blue) [Eq.~\ref{HFapp}] in a system with $L=60$ unit cells, for the $B=4$, $a_0=1$ EB state with support in $x\in [0,30]$. Evidently, there is more scrambling and hence faster oscillations when more of the EB state falls outside the potential well. (b) 2-point function evolution for different potential wall slopes $\alpha=1,2,10$ (light green, teal, blue) [Eq.~\ref{HFapp}] for a random state with uncorrelated uniformly distributed site amplitudes. A more well-defined potential well leads to more regular oscillations.}
\label{figa:corrEx2}
\end{figure}

The extent of decay depends on the Hamiltonian $H_F$'s energy profile, as color coded in Fig.~\ref{figa:corrEx}d. In this work, we have considered $H_F$ of the form
\begin{equation}
H_F=\sum_x E(x)|x\rangle\langle x|=\sum_x\frac{E_0}{1+e^{\alpha(x_0-x)}}|x\rangle\langle x|
\label{HFapp}
\end{equation}
such that it it creates a potential well with soft walls at $x=x_0$, with larger $\alpha$ leading to a steeper well. In Figs.~\ref{figa:corrEx} and \ref{figa:corrEx2}a, we have respectively investigated the effects of tuning $\alpha$ and $x_0$. In Fig.~\ref{figa:corrEx}, where $x_0$ is fixed at $L/2$, the 2-point function evolution is compared between (a) a $B=4$ EB state ``bounded'' at $x=L/2$, i.e. defined by $R= \sum_x\theta(L/2-x)|x\rangle\langle x|$, (b) an uniform state $\varphi(x)=\theta(L/2-x)$ between $x=0$ and $x=L/2$, and (c) an uniform  state $\varphi(x)=\theta(L/4-x)$ between $x=0$ and $x=L/4$. The various curves correspond to $H_F$ profiles labeled in Fig.~\ref{figa:corrEx}d by the same color. For the EB state (a), the 2-point function fluctuates significantly for very gentle potential wells due to the nonzero tail of the nonzero energy within $[0,L/2]$. This nonzero energy tail scrambles the state $\varphi(x)$, and is more pronounced when the well boundary is softer, as evidenced in (b) and (c). While a relatively hard $L/2$ boundary with $\alpha=100$ (purple) leads to slow oscillations for (b) $\varphi(x)=\theta(L/2-x)$ and non-existent fluctuations for (c) $\varphi(x)=\theta(L/4-x)$, very soft boundaries with $\alpha=1$ (blue) and $\alpha=2$ (orange) lead to rapid initial decay, followed by partially recurrent oscillations. Over time, these oscillations completely lose their coherence and descend into increasingly random oscillations.

The effects of tuning the potential wall position $x_0$ are presented in Fig.~\ref{figa:corrEx2}a for $\varphi(x)$ being the same $B=4$ EB state and potential slope $\alpha=10$. For $x_0/L=5/12<1/2$ (red), the nonzero energy tail within the support of the EB state ($0\leq x \leq L/2$) scrambles the state, leading to oscillations. This oscillations become slower as the tail moves away, as plotted for $x_0/L=1/2$ (magenta) and $x_0/L=7/12$ (blue). 

It is also illuminating to investigate the effects of the scrambling by $H_F$ on $\varphi(x)$ being a random state with uncorrelated uniformly distributed amplitudes at each site. Shown in Fig.~\ref{figa:corrEx2}b are the 2-point functions for $x_0=L/2$ and different slopes $\alpha=1,2,10$ (light green, teal, blue). As the slope increases and the potential well becomes more well-defined, scrambling due to time-evolution becomes confined to the part of the wavefunction outside the well, leading to an approximately two-component effective system. This manifest as regular oscillations in the 2-point function.

\section{Quantum entanglement and EB state phenomena}

\subsection{General results}
In the general many-body context, a system consists of occupied single-body states $\{|\chi_{\scaleto{\xi}{4pt}}\rangle\}$ that collectively form a many-body state $|\Psi\rangle$. The latter defines the density matrix via $\rho = |\Psi\rangle\langle \Psi|$, where bras are understood to be the biorthogonal conjugate of the kets with respect to the non-Hermitian Hamiltonian (unlike in the main text, they are not indicated with a tilde for brevity). Entanglement information is contained in the reduced density matrix $\rho_{\scaleto{\mathcal{R}}{4pt}}=\text{Tr}_{\scaleto{\mathcal{R}^c}{4pt}}\rho$, which is obtained from $\rho$ by tracing out degrees of freedom in $\mathcal{R}^c$, where $\mathcal{R}^c$ is the complement of a prescribed subregion $\mathcal{R}$.

To put $\rho_{\scaleto{\mathcal{R}}{4pt}}$ in a palatable form, we would like to rotate the single-body basis into a new basis $\{|\chi'_{\scaleto{\xi}{4pt}}\rangle\}$ such that it consists of basis states that are either wholly contained in $\mathcal{R}$ or $\mathcal{R}^c$. To do so, we first explicitly write down the many-body state as
\begin{equation}
|\Psi\rangle = \bigotimes_{\xi\in \text{occ}} |\chi_{\scaleto{\xi}{4pt}}\rangle = \bigotimes_{\xi} P|\chi_{\scaleto{\xi}{4pt}}\rangle=\bigotimes_{\xi} (RP|\chi_{\scaleto{\xi}{4pt}}\rangle +R^cP|\chi_{\scaleto{\xi}{4pt}}\rangle ),
\end{equation}
where $R$ and $R^c=\mathbb{I}-R$ are projectors onto regions $\mathcal{R}$ and $\mathcal{R}^c$ respectively, and the occupied subspace projector $P=\sum_{\xi\in\text{occ}}|\chi_{\scaleto{\xi}{4pt}}\rangle\langle\chi_{\scaleto{\xi}{4pt}}|$ ensures that only occupied states are included. Importantly, we would like the two terms $RP|\chi_{\scaleto{\xi}{4pt}}\rangle$ and $R^cP|\chi_{\scaleto{\xi}{4pt}}\rangle $ on the right to be each proportional to a basis state of $\{|\chi'_{\scaleto{\xi}{4pt}}\rangle\}$. Suppose $|\chi'_{\scaleto{\xi}{4pt}}\rangle\propto RP|\chi_{\scaleto{\xi}{4pt}}\rangle$. Then 
\begin{equation}
\langle\chi'_{\scaleto{\mu}{4pt}}|\chi'_{\scaleto{\xi}{4pt}}\rangle \propto \langle\chi_{\scaleto{\mu}{4pt}}|(PR)(RP)|\chi_{\scaleto{\xi}{4pt}}\rangle=\langle\chi_{\scaleto{\mu}{4pt}}|PRP|\chi_{\scaleto{\xi}{4pt}}\rangle = r_\xi\delta_{\mu\xi}
\label{overlap}
\end{equation}
if we let $|\chi_{\scaleto{\xi}{4pt}}\rangle$ be an eigenstate of $\tilde R=PRP$ with eigenvalue $r_\xi$, i.e. $\tilde R |\chi_{\scaleto{\xi}{4pt}}\rangle= r_\xi |\chi_{\scaleto{\xi}{4pt}}\rangle$. In Ref.~\cite{chang2020entanglement}, $\langle \psi_\alpha |\tilde R|\psi_\beta\rangle$ was interpreted as the overlap matrix between arbitrary states $|\psi_\alpha\rangle$ and $|\psi_\beta\rangle$. With Eq.~\ref{overlap}, we can construct the requisite new orthonormal basis $\{|\chi'_{\scaleto{\xi}{4pt}}\rangle\}$ in $\mathcal{R}$ via
\begin{equation}
|\chi'_{\scaleto{\xi}{4pt}}\rangle=\frac1{\sqrt{r_\xi}}RP|\chi_{\scaleto{\xi}{4pt}}\rangle.
\label{newbasis}
\end{equation}
Since an eigenvalue $r_\xi$ always occurs together with $1-r_\xi$, we also have the following basis states that are wholly contained in $\mathcal{R}^c$:
\begin{equation}
|\chi'^c_{\scaleto{\xi}{4pt}}\rangle=\frac1{\sqrt{1-r_\xi}}R^cP|\chi_{\scaleto{\xi}{4pt}}\rangle.
\end{equation}
Hence the many-body state is expressed as
\begin{equation}
|\Psi\rangle =\bigotimes_{\xi} (\sqrt{r_\xi}|\chi'_{\scaleto{\xi}{4pt}}\rangle +\sqrt{1-r_\xi}|\chi'^c_{\scaleto{\xi}{4pt}}\rangle ).
\end{equation}
%which separates nicely into contributions from $R$ and $R^c$.
In this form, the contributions from $\mathcal{R}$ and $\mathcal{R}^c$ are separated, and the trace over $\mathcal{R}^c$ can be easily done to obtain $\rho_{\scaleto{\mathcal{R}}{4pt}}=\text{Tr}_{\scaleto{\mathcal{R}^c}{4pt}}\rho$:
\begin{eqnarray}
\rho_{\scaleto{\mathcal{R}}{4pt}}=\text{Tr}_{\scaleto{\mathcal{R}^c}{4pt}}|\Psi\rangle\langle \Psi|&=&\text{Tr}_{\scaleto{\mathcal{R}^c}{4pt}}\bigotimes_{\xi,\mu}(\sqrt{r_\xi}|\chi'_{\scaleto{\xi}{4pt}}\rangle +\sqrt{1-r_\xi}|\chi'^c_{\scaleto{\xi}{4pt}}\rangle )(\sqrt{r_\mu}\langle\chi'_{\scaleto{\mu}{4pt}}| +\sqrt{1-r_\mu}\langle\chi'^c_{\scaleto{\mu}{4pt}}| )\notag\\
&=&\text{Tr}_{\scaleto{\mathcal{R}^c}{4pt}}\bigotimes_{\xi,\mu}(\sqrt{r_\xi r_\mu}|\chi'_{\scaleto{\xi}{4pt}},0\rangle\langle \chi'_{\scaleto{\mu}{4pt}},0|+\sqrt{r_\xi (1-r_\mu)}|\chi'_{\scaleto{\xi}{4pt}},0\rangle\langle 0,\chi'^c_{\scaleto{\mu}{4pt}}|+\sqrt{r_\mu (1-r_\xi)}|0,\chi'^c_{\scaleto{\xi}{4pt}}\rangle\langle \chi'_{\scaleto{\mu}{4pt}},0|\notag\\
&&\qquad+\sqrt{(1-r_\xi)(1-r_\mu)}|0,\chi'^c_{\scaleto{\xi}{4pt}}\rangle\langle 0,\chi'^c_{\scaleto{\mu}{4pt}}\rangle|)\notag\\
&=&\sum_\gamma \langle 0,\chi'^c_{\scaleto{\gamma}{4pt}}|\bigotimes_{\xi,\mu}(\sqrt{r_\xi r_\mu}|\chi'_{\scaleto{\xi}{4pt}},0\rangle\langle \chi'_{\scaleto{\mu}{4pt}},0|+\sqrt{r_\xi (1-r_\mu)}|\chi'_{\scaleto{\xi}{4pt}},0\rangle\langle 0,\chi'^c_{\scaleto{\mu}{4pt}}|+\sqrt{r_\mu (1-r_\xi)}|0,\chi'^c_{\scaleto{\xi}{4pt}}\rangle\langle \chi'_{\scaleto{\mu}{4pt}},0|\notag\\
&&\qquad+\sqrt{(1-r_\xi)(1-r_\mu)}|0,\chi'^c_{\scaleto{\xi}{4pt}}\rangle\langle 0,\chi'^c_{\scaleto{\mu}{4pt}}\rangle|)|0,\chi'^c_{\scaleto{\gamma}{4pt}}\rangle\notag\\
&=&\bigotimes_\xi \left(r_\xi|\chi'_{\scaleto{\xi}{4pt}}\rangle\langle\chi'_{\scaleto{\xi}{4pt}}|+(1-r_\xi)|0\rangle\langle 0|\right).
\label{rhoR}
\end{eqnarray}
In the second and third lines above, we have used the notation $|a,b\rangle = |a\rangle_{\mathcal{R}}|b\rangle_{\mathcal{R}^c}$ to indicate whether the state belongs to $\mathcal{R}$ or $\mathcal{R}^c$. Note that $\text{Tr}_{\scaleto{\mathcal{R}^c}{4pt}} |a,0\rangle\langle a,0| = \sum_b \langle 0,b|a,0\rangle\langle a,0| 0,b\rangle=|a\rangle\langle a|$.

%  To see why, we label $r_\xi$ as the eigenvalue corresponding to $|\chi_{\scaleto{\xi}{4pt}}\rangle$, i.e. $\tilde R|\chi_{\scaleto{\xi}{4pt}}\rangle=r_\xi |\chi_{\scaleto{\xi}{4pt}}\rangle$. Then, since $|\chi_{\scaleto{\xi}{4pt}}\rangle$ is by definition also an eigenvector of $P$, we have

The EE for region $\mathcal{R}$ can be obtained via
\begin{equation}
S_\mathcal{R}= -\text{Tr}\rho_{\scaleto{\mathcal{R}}{4pt}}\log\rho_{\scaleto{\mathcal{R}}{4pt}}=-\sum_\xi[r_\xi\log r_\xi+(1-r_\xi)\log(1-r_\xi)].
\label{Sr}
\end{equation}

\subsection{Relation to the 2-site propagator and EB states}

For the case of free fermions, the anticommutation algebra makes it possible to express the reduced density matrix as $\rho_{\scaleto{\mathcal{R}}{4pt}}=e^{-H_E}/(\mathbb{I}+e^{-H_E})=(\mathbb{I}+e^{H_E})^{-1}$, where $H_E$ is the \emph{single-particle} entanglement Hamiltonian. In this case, $H_E=\log(\bar P^{-1}-\mathbb{I})$ where $\bar P = RPR$ is the (single-particle) correlation matrix. The real-space elements of $\bar P$ constitute the 2-site propagator, as explained in the main text. In particular, it was $\bar P$ which hosts EB states as eigenstates, even though it was the eigenstates of $\tilde R$ that enter the Schmidt decomposition of the entanglement cut as well as the reduced density matrix [Eq.~\ref{rhoR}]. 

Let us now elucidate the relationship between $\bar P =RPR$ and $\tilde R = PRP$. Given an arbitrary eigenstate $|\phi_\xi\rangle$ of $\bar P$, i.e. $\bar P|\phi_\xi\rangle=p_\xi |\phi_\xi\rangle $, we have 
\begin{equation}
\tilde R(P|\phi_\xi\rangle)=PRP^2|\phi_\xi\rangle=PRP|\phi_\xi\rangle=(PRP)R|\phi_\xi\rangle=p_\xi (P|\phi_\xi\rangle)
\end{equation}
where we have used $R|\phi_\xi\rangle=|\phi_\xi\rangle$. In other words, $|\chi_{\scaleto{\xi}{4pt}}\rangle=P|\phi_\xi\rangle/\sqrt{\langle \phi_\xi|P|\phi_\xi\rangle}=\frac1{\sqrt{p_\xi}}P|\phi_\xi\rangle$ is a normalized eigenstate of $\tilde R$ with identical eigenvalue $r_\xi=p_\xi$. Immediately, this means that we can also express the EE in terms of the eigenvalues $p_\xi$ of $\bar P$ i.e. 
\begin{equation}
S_\mathcal{R}= -\sum_\xi[p_\xi\log p_\xi+(1-p_\xi)\log(1-p_\xi)]
\label{Sp}
\end{equation}
which is identical to Eq.~\ref{Sr} except that $r_\xi$ is replaced by $p_\xi$. However, the reduced density matrix takes on a slightly simpler form when expressed in terms of the eigenstates $|\phi_\xi\rangle$ of $\bar P$. Substituting $|\chi_{\scaleto{\xi}{4pt}}\rangle=\frac1{\sqrt{p_\xi}}P|\phi_\xi\rangle$ into Eq.~\ref{newbasis}, we have 
\begin{equation}
|\chi'_{\scaleto{\xi}{4pt}}\rangle=\frac1{\sqrt{r_\xi}}RP|\chi_{\scaleto{\xi}{4pt}}\rangle=\frac1{\sqrt{p_\xi r_\xi}}RP^2|\phi_\xi\rangle=\frac1{\sqrt{p_\xi r_\xi}}RPR|\phi_\xi\rangle=\frac{p_\xi}{\sqrt{p_\xi r_\xi}}|\phi_\xi\rangle=|\phi_\xi\rangle,
\end{equation}
such that 
\begin{equation}
\rho_{\scaleto{\mathcal{R}}{4pt}}=\bigotimes_\xi \left(p_\xi|\phi_\xi\rangle\langle\phi_\xi|+(1-p_\xi)|0\rangle\langle 0|\right)
\label{rhoR2}
\end{equation}
which is expressed in terms of the \emph{unrotated} eigenstates of $\bar P$. In particular, for a system containing an EP with $B\geq 2$, at least one of the $|\phi_\xi\rangle$ will be an EB eigenstate $|\phi\rangle$ with eigenvalue $p_\xi=p\notin [0,1]$. %the weight of the $|\phi_\xi\rangle\langle\phi_\xi|$ term will be asymmetrically enhanced 
In the large $p$ limit, each EB Schmidt factor $\left(p_\xi|\phi_\xi\rangle\langle\phi_\xi|+(1-p_\xi)|0\rangle\langle 0|\right)$ approximates $p(|\phi\rangle\langle\phi|-|0\rangle\langle 0|)$, which is the enigmatic consequence of the singular asymmetric propagator.

\section{EB phenomena for specific models}

In this last section, we present a table of various EP lattice Hamiltonians with some of their quantities that are relevant to EB phenomena.

\begin{table*}[h]
		\begin{tabular}{ |c |c| c | c |c| c | c |c|c|c|}
	\hline
		\bf Model  & $B$  & $a_0$ &$b_0$ & $\gamma(k)$ & $U_x\sim$& $\lambda(L)\sim$ & $S(L)\sim$ \\ \hline
			  $(2(1-\cos k)+a_0)\sigma_x+a_0\sigma_+$ & 2& $a_0>0$ & 1 & 0 & $2\sqrt{\frac{a_0}{b_0}}\log\frac{L}{x}$ & $\log L$ & $-\frac{2}{3}\log L$  \\
				 $(2(1-\cos k)+a_0)^2\sigma_x+a_0\sigma_+$ & 4& $a_0>0$ & 1& 0 &$\frac{2}{\pi}\sqrt{\frac{a_0}{b_0}}(L-2x)$& $L$ & $-2\log L$  \\
								 $(2(1-\cos k)+a_0)^3\sigma_x+a_0\sigma_+$ & 6& $a_0>0$ & 1& 0 &$\left(\frac{2}{\pi}\right)^2\sqrt{\frac{a_0}{b_0}}\left(L^2-\frac{\pi^2x^2}{2}\right)$& $L^2$ & $-4\log L$  \\
								$(2(1-\cos k)+a_0)^6\sigma_x+a_0\sigma_+$ & 12& $a_0>0$ & 1& 0 &$2\sqrt{\frac{a_0}{b_0}}\left(\frac{L}{\pi}\right)^{5}\left(2-\frac{\pi^2x^2}{L^2}\right)$& $L^5$ & $-18\log L$  \\
				$(v-w\cos k)\sigma_x+\gamma_0\sin k\,\sigma_y+i(v-w)\sigma_z$ & 2& $2(v-w)$&$\frac{w}{2}$ & $\gamma_0\sin k$ & depends on $\gamma_0$& depends on $\gamma_0$ & $-\frac{c'(\gamma_0)}{3}\log L$  \\
																$2(1-\cos k)\sigma_x+\sin k\,\sigma_y$ & 2& $0$&$\frac{1}{2}$ & $\sin k$ &  1/x & $\approx 0$ & $\frac{1}{3}\log L$  \\
							\hline
						\end{tabular}
			\caption{ Various critical lattice models and the forms of their $a_0,b_0$ and $\gamma(k)$. $U_x=-2\langle c^\dagger_{x,+}c_{0,-}\rangle$ is the propagator from the $-$ to $+$ sublattice across $x$ unit cells. $\lambda(L)$ is the eigenvalue of $\Lambda=4(\bar P^2-\bar P)$ corresponding to the EB eigenstate, whose $\lambda(L)$ depends on $L$ and which can be very large. $S(L)$ is the entanglement entropy scaling behavior at half-filling, comprising contributions from EB states as well as all other eigenstates. The first four models are of the prototypical form introduced in the text, with different powers $B$. The fourth model however has an anomalously large negative logarithmic EE scaling coefficient of $-18$, which arises due to the secondary EB states from its large $B$. They cannot be predicted from $\lambda(L)$, which only pertains to the (primary) EB states. The following (fifth) model is introduced in the final section of the main text and has a nontrivial $\gamma(k)$. It possesses complicated $U_x$ and $\lambda(L)$ dependencies, with ``central charge''	$c'(\gamma_0)$ given in the inset of Fig 3a. The last model is a Hermitian critical point with no EB states to speak of ($\lambda(L)\approx 0$), and exhibits ordinary $\frac1{3}\log L$ EE scaling.}
		\label{table:Cap}
	\end{table*}

\end{document}